\documentclass[aps,prd,twocolumn,preprintnumbers,superscriptaddress,dblfloatfix,nofootinbib]{revtex4-2}

\usepackage{graphicx}
\usepackage{dcolumn}
\usepackage{bm}
\usepackage{xcolor}
\usepackage{url}
\usepackage{amsmath}
\usepackage{comment}
\usepackage[normalem]{ulem}

\newcommand{\bfk}{\mbox{\boldmath$k$}}

\newcommand{\bfn}{\mbox{\boldmath$n$}}

\newcommand{\hiGpc}{h^{-1}\mathrm{Gpc}}
\newcommand{\hiMpc}{h^{-1}\mathrm{Mpc}}
\newcommand{\hikpc}{h^{-1}\mathrm{kpc}}

\newcommand{\hMpci}{h\,\mathrm{Mpc}^{-1}}

\newcommand{\Mvir}{M_\mathrm{vir}}

\newcommand{\kmax}{k_\mathrm{max}}

\def\be{\begin{equation}}
\def\ee{\end{equation}}
\def\k{\boldsymbol{k}}
\def\p{\boldsymbol{p}}
\def\q{\boldsymbol{q}}

\def\z{\boldsymbol{z}}

\begin{document}
\title{Blinded
challenge
for precision cosmology with large-scale structure: results from
effective field theory for the redshift-space galaxy power spectrum}

\author{Takahiro Nishimichi}
\affiliation{Center for Gravitational Physics, \\ Yukawa Institute for Theoretical Physics, Kyoto University, Kyoto 606-8502, Japan}
\affiliation{Kavli Institute for the Physics and Mathematics of the Universe (WPI), UTIAS \\The University of Tokyo, Kashiwa, Chiba 277-8583, Japan}

\author{Guido D'Amico}
\affiliation{Stanford Institute for Theoretical Physics, Physics Department,\\ Stanford University, Stanford, CA 94306}
\affiliation{Dipartimento di SMFI dell' Universita' di Parma \& INFN Gruppo Collegato di Parma, Parma, Italy}

\author{Mikhail M. Ivanov}

\affiliation{Center for Cosmology and Particle Physics, Department of Physics, New York University, New York, NY 10003, USA}
\affiliation{Institute for Nuclear Research of the
Russian Academy of Sciences, \\
60th October Anniversary Prospect, 7a, 117312
Moscow, Russia
}

\author{Leonardo Senatore}
\affiliation{Stanford Institute for Theoretical Physics, Physics Department,\\ Stanford University, Stanford, CA 94306}
\affiliation{Kavli Institute for Particle Astrophysics and Cosmology,\\
 SLAC and Stanford University, Menlo Park, CA 94025}

\author{Marko Simonovi\'c}
\affiliation{Theoretical Physics Department, CERN,  \\ 1 Esplanade des Particules, Geneva 23, CH-1211, Switzerland}

\author{Masahiro Takada}
\affiliation{Kavli Institute for the Physics and Mathematics of the Universe (WPI), UTIAS \\The University of Tokyo, Kashiwa, Chiba 277-8583, Japan}

\author{Matias Zaldarriaga}
\affiliation{School of Natural Sciences, Institute for Advanced Study, \\ 1 Einstein Drive, Princeton, NJ 08540, USA}

\author{Pierre Zhang}
\affiliation{Department of Astronomy, School of Physical Sciences, University of Science and Technology of China, Hefei, Anhui 230026, China}
\affiliation{CAS Key Laboratory for Research in Galaxies and Cosmology, University of Science and Technology of China, Hefei, Anhui 230026, China}
\affiliation{School of Astronomy and Space Science, University of Science and Technology of China, Hefei, Anhui 230026, China}

%
\date{\today}
\begin{abstract}
An accurate theoretical template for the galaxy power spectrum is
a key for the success of ongoing and future spectroscopic surveys.
We examine to what extent the Effective Field Theory of Large Scale Structure
is able to provide such a template and correctly estimate cosmological parameters.
To that end, we initiate a blinded challenge to infer cosmological parameters
from the redshift-space power spectrum of high-resolution mock catalogs
mimicking the BOSS galaxy sample but covering a hundred times larger
cumulative
volume.
This gigantic simulation volume allows us to separate systematic bias due to theoretical modeling
from the statistical error due to sample variance.
The challenge task was to measure three unknown input parameters used in the simulation:
the Hubble constant, the matter density fraction, and the clustering amplitude.
We present analyses done by two independent teams,
who have fitted the
mock
simulation
data generated by yet another
independent
group.
This allows us to avoid any
confirmation
bias by analyzers and pin down possible tuning of the specific EFT implementations.
Both independent teams have recovered the true values of the input parameters within sub-percent statistical
errors corresponding to
the total simulation volume.
\end{abstract}

\preprint{YITP-20-25}
\preprint{INR-TH-2020-009}
\preprint{CERN-TH-2020-040}
\preprint{IPMU20-0025}
\maketitle

\section{Introduction}
\label{sec:intro}
Modern cosmology is getting more and more mature as accumulating observational data are available for us. We have however a fundamental lack of understanding of the physical nature of the dark components introduced to explain the dominant source of gravity that gathers material to form rich structures in the late universe (dark matter) as well as the accelerating cosmic expansion (dark energy), together filling the majority of the cosmological energy budget. Aiming at having further insights on those substances, growing number of large-scale observational programs are ongoing and planned \citep[e.g.][]{2014PASJ...66R...1T,laureijs2011,2009arXiv0912.0201L,DESI}.

Of crucial importance from the theoretical point of view is our ability to prepare an accurate model template with which one can confront such observational data for their interpretation. Since a larger survey means a smaller statistical error, the relative contribution from the systematic error arising from the inaccuracy of the template should be more important. Given the gigantic area coverage and depth of ambitious future programs, there is the necessity to come up with a really accurate theoretical framework to predict the observed large scale structure to
attain their full potential to infer the underlying theory governing the universe.

One of the most difficult aspect of the large-scale structure prediction is the complicated relation between the matter density fluctuations dominated by invisible dark matter and visible structures such as galaxies \cite{kaiser84}. The so-called galaxy bias cannot be predicted from first principles, unless one can model all the baryonic effects relevant for the formation and evolution of galaxies.
While hydrodynamical simulations might be one way to proceed, the vastly large dynamical range, kpc to Gpc in length scale, is a big obstacle. Typically, one comes up with empirical subgrid models and calibrate them against observed statistics of galaxies \citep[see e.g., ][for recent attempts]{2014Natur.509..177V,2014MNRAS.445..175G,Vogelsberger_2014,2015MNRAS.450.1937C,2015MNRAS.446..521S,2014MNRAS.444.1453D,2018MNRAS.475..676S,2019ComAC...6....2N}.

Alternatively, one can formulate the statistical properties of galaxies on large scales
via
a perturbative expansion
in which poorly known galaxy physics is parameterized
by a set of effective bias
operators.
The strength of these operators is
controlled by free
coefficients,
which should be treated
as nuisance parameters.
The recently developed Effective Field Theory of Large Scale Structure (EFTofLSS) provides a systematic way to derive all possible operators
and corresponding bias coefficients
that are
allowed by symmetry
\citep{McDonald:2009dh,baumann12,carrasco12,Assassi:2014fva,Senatore:2014via,Senatore:2014eva,Senatore:2014vja,Lewandowski:2014rca,Lewandowski:2015ziq} \citep[also see][for a review]{Desjacques18}.
Since this approach, in principle, does not assume any specific model of galaxy formation, it provides us
with a conservative theoretical model for
the galaxy density and velocity
fields on large scales.
The generality of the effective field theory approach comes at
the price of having to marginalize over many free coefficients,
which can compromise cosmological constraints.
These constraints can become weaker compared to other theoretical templates in which a specific bias prescription is employed, such as halo model approaches. The detailed balance between the robustness and the tightness of the cosmological constraints has been addressed in recent studies \citep[e.g.,][]{Hand:2017ilm,2020PhRvD.101b3510K,Osato_2019}.

There are several non-trivial choices
behind the application
of the EFT to the data.
First, one should determine the
wavenumber up to which
the EFT calculation
up to a chosen perturbative order is
reliable.
This data cut should be carefully
tested to avoid biased parameter estimates.
Then, one has to decide
how many nuisance parameters to keep
in the fit (there are about 10 at the one-loop order) and what priors
to use.
Indeed, at the
power spectrum level many
EFT operators are degenerate
among each other.
Thus, one has to accurately
determine their principal
components to make the cosmological
analysis efficient.
All these subtleties should be examined and validated in a transparent manner to convince the community of the robustness of the EFT approach.

To that end, in this paper, we conduct a first \textit{blind} test of EFTofLSS for clustering of galaxies in redshift space. Two independent groups, which will be referred to as ``West Coast'' (D'Amico, Senatore and Zhang) and ``East Coast'' (Ivanov, Simonovi\'c and Zaldarriaga), have analyzed the mock data generated by yet another group
(Nishimichi and Takada,
simply ``Japan Team''  hereafter).
In this process, the
true cosmological parameters used to
generate the simulation mock data were known only to the Japan Team.
The two analyzing teams have participated
in the challenge on the condition
that the results would be published regardless of the outcome,
and the pipelines could not be
modified after unblinding.
We present these results in our paper in the original form.
To complement the result of the blinded analysis and to get more insight on the origin of the cosmological information, we briefly discuss post-unblinding analyses.

The layout of this paper is as follows. We first describe the design of our mock challenge program in Sec.~\ref{sec:challenges}. We then specify the mock simulations in Sec.~\ref{sec:mock}. The theoretical template and the method to conduct parameter inference are explained in Sec.~\ref{sec:theory}. Then the results of the blinded analysis are summarized in Sec.~\ref{sec:res_blind}.
We conclude this study in Sec.~\ref{sec:conclusion}.

\section{Design of Blinded Cosmology Challenge}
\label{sec:challenges}

Throughout this paper, we consider a flat $\Lambda$CDM cosmology. This is motivated by the recently claimed tension in the values of the Hubble parameter, one from local measurements such as the distance ladder, and the other from the Cosmic Microwave Background (CMB) assuming a flat $\Lambda$CDM model \citep[see][and references therein]{2019NatRP...2...10R}. In such a situation, a robust measurement from other independent observable channels would be important, and indeed, the galaxy clustering, when the full shape information of its spectra is analysed, has been shown to serve as such a probe \citep{DAmico:2019fhj,Ivanov:2019pdj,Colas:2019ret,2020A&A...633L..10T}. Also important might be a similar, but a weaker tension in the amplitude of the density fluctuations in the current universe
\citep{Hildebrandt17,2018PhRvD..98d3526A,Hikage19}.
This is known to be degenerate with the matter density parameter from the late-time observables.
We wish to demonstrate through the challenge the current status of the use of galaxy clustering in particular with an EFT approach to describe the nonlinear nature of the cosmological large scale structure.

\subsection{Cosmological parameters}
\label{subsec:cosmo}

To assess the reliability of the galaxy-clustering analyses within the flat $\Lambda$CDM model, three cosmological parameters, $\ln(10^{10}A_\mathrm{s})$, $\Omega_\mathrm{m}$ and $H_0$, are randomly drawn from independent normal distributions. These parameters are the logarithm of the amplitude of the primordial power spectrum at $k_0=0.05\,\mathrm{Mpc}^{-1}$, the matter density parameter at present and the current Hubble expansion rate in km/s/Mpc, respectively. While the mean values of the normal distributions are set to be the best-fit values determined by Planck satellite \cite{planck-collaboration:2015fj}, we consider the standard deviation four times larger than the same experiment to test the validity of the model in a broader parameter space. While all of the information above is shared among all the collaborators, the three random numbers drawn were kept only within the Japan Team until we finally unblinded them.

On the other hand, we fix the baryon fraction, $f_\mathrm{b} = 0.1571$ and the spectral index $n_\mathrm{s}=0.9649$. These values are shared with the two US teams. In typical current large-scale structure survey analyses, these two parameters are not very well determined due to the weak sensitivity of the target galaxy observables unless one adds priors motivated by CMB observations and/or big-bang nucleosynthesis, while it would be possible to constrain them from futuristic galaxy surveys. Therefore, letting the US analysis teams know the exact values of them loosely corresponds to adding CMB priors \footnote{It is not trivial how one can best arrange a challenge where external prior information is added. To keep the analysis fully blinded, the Japan Team decided not to give any prior information to the analysis teams for the challenge presented in this paper.}. Further, for simplicity and to avoid the complication to deal with massive neutrinos both in theory and in simulations, we set the neutrino masses to be exactly zero. Under the above settings, the linear matter-density transfer function is computed using the public Boltzmann solver \textsc{CAMB} \cite{camb}. The parameter file passed to this code by the Japan Team is provided to the US teams after
the values of
$\omega_\mathrm{b}$, $\omega_\mathrm{c}$, $H_0$ and
$A_\mathrm{s}$
are erased.

The main goal of the challenge is to infer the three cosmological parameters $A_\mathrm{s}$, $\Omega_\mathrm{m}$ and $H_0$. It was agreed among all the teams that, once these cosmological parameters are unblinded, the
results reported by the time may not be modified any more.

\subsection{Target observables}
\label{subsec:obs}
We focus on the galaxy clustering in redshift space in the initial challenge presented in this paper. More specifically, we work in Fourier space and analyse the multipole moments of the galaxy power spectrum. This includes physical and observational effects such as the Baryon Acoustic Oscillation (BAOs; \cite{1970ApJ...162..815P,1970Ap&SS...7....3S,1984ApJ...285L..45B,1987MNRAS.226..655B,1989ApJS...71....1H}), redshift-space distortions (RSD; \cite{jackson72,kaiser87}) and the Alcock-Paczynski (AP; \cite{alcock79}) effect, where the AP is induced artificially by distorting the simulation boxes (see the next section for further detail). On top of these distinctive features, the mock data should contain the cosmological information through the overall shape of the power spectra, which might be hindered by the presence of various nonlinear effects. The aim of this challenge is to assess how robustly one may extract the fundamental cosmological parameters within the flat $\Lambda$CDM framework.

The Japan Team constructs mock galaxy catalogs and measures the multipole moments of the power spectra. To discriminate the \textit{systematic} error from the statistical error, this experiment is done in huge simulation volumes much larger than the current surveys. The galaxy catalogs are constructed to roughly mimic the CMASS and the LOWZ catalog from the
12th
Data Release
of Sloan Digital Sky Survey \citep[Ref.][hereafter SDSS DR12]{2015ApJS..219...12A}. The details of these simulations will follow in the next section. Since the galaxy bias is formulated to be as general as possible in the EFT, based only on symmetry considerations without assuming any specific model with which galaxies are defined, the detail of the mock galaxies would not give a significant impact to the blinded analysis as long as one sticks to an EFT approach. However, other approaches such as the halo model would be directly impacted by the piece of information on the exact procedure with which the mock galaxies are distributed within the simulation volume. Therefore, any further information on the mock galaxies detailed in the next section was not provided to the US teams before unblinding.

For completeness, the set of mock data as well as the information on the simulations provided to the US teams are summarised at a dedicated website (\url{http://www2.yukawa.kyoto-u.ac.jp/~takahiro.nishimichi/data/PTchallenge/}). All the data and the information were shared through this website. Interested readers may download the same set of data and participate in the blinded challenge by analysing the data using their own theoretical template, as the exact cosmological parameter values are not exactly shown in this paper nor on the website.

\section{Generating mock redshift-space power spectra of BOSS-like galaxies}
\label{sec:mock}

The Japan Team works on the construction of mock galaxy catalogs and measurement of the power spectra. The settings of the numerical simulations, the prescription for the mock galaxies and the analysis methods to determine their statistics are described in this section.

\subsection{Specification of simulations}
\label{subsec:simu}
We follow the gravitational dynamics of ten random realizations of the matter density field expressed by $3,072^3$ mass elements sampled in comoving periodic cubes with the side length $L = 3,840\,\hiMpc$. The total volume,  $566\,(\hiGpc)^{3}$, is about
a hundred times that of the CMASS and LOWZ sample from SDSS BOSS DR12, which together have a volume coverage of
$5.7 \, (\hiGpc)^3$ \citep{2013AJ....145...10D}.
The large volume of our simulations allows us to determine the statistics of the mock galaxies very precisely with little sample-variance
error. Therefore, we can conduct a fairly stringent test of the systematic error due to an imperfect modeling of the target statistics.

The initial conditions are generated with a code developed in \cite{nishimichi09} and then parallelized in \cite{Valageas11a} based on the second-order Lagrangian Perturbation Theory (2LPT; \cite{scoccimarro98,crocce06a}). Following the result presented in \cite{DQ1}, the starting redshift of the simulations are set at $z=29$ to roughly optimize the total systematic error arising from the artificial growing mode due to the grid pre-initial condition \cite{Marcos06,Joyce07,Garrison16} and the truncation of the LPT at the second order given the mean inter-particle distance of the simulations. We prepare ten independent random realizations, each of which is then evolved by a public Tree-Particle Mesh code \textsc{Gadget2}~\cite{gadget2} with $6,144^3$ grid points for fast Fourier transform (FFT) and the tree softening length of $62.5\,\hikpc$. The other simulation parameters to control the force accuracy as well as the time-stepping criteria are the same as in \cite{DQ1}. We store the particle snapshots at $z=3$, $2$, $1$, $0.61$, $0.51$ and $0.38$. We populate galaxies to the lowest three redshifts, and conventionally call the catalogs as CMASS2 ($z=0.61$), CMASS1 ($z=0.51$) and LOWZ ($z=0.38$) in what follows.

\subsection{Mock galaxy identification}
\label{subsec:galaxies}
After obtaining the particle snapshots, we run the
\textsc{Rockstar} halo finder~\cite{Behroozi:2013}, which is based on the six dimensional phase space friends-of-friends algorithm. This code identifies not only isolated ``central'' halos but also ``satellite'' halos existing as substructures of more massive halos without any distinction at first in the primary output files. For simplicity, we treat each of them irrespectively of whether it is a central or a satellite halo and populate a galaxy only according to the virial mass assigned by \textsc{Rockstar}. We impose a soft cutoff to the virial mass to select massive halos to populate galaxies randomly with the probability
\begin{eqnarray}
P(\Mvir) = \frac{1}{2}\left[1+\mathrm{erf}\left(\frac{\log_{10}\Mvir-\log_{10}M_\mathrm{min}}{\sigma_{\log_{10}M}}\right)\right],\label{eq:HOD}
\end{eqnarray}
where ${\rm erf}(x)$ is the error function.
We have two parameters, $\log_{10}M_\mathrm{min}$ and $\sigma_{\log_{10}M}$, which determine the typical minimum mass and the profile of the soft mass cutoff, respectively.
We set $\log_{10} M_\mathrm{min} = 13.08$, $12.97$ and $12.95$ for LOWZ, CMASS1 and CMASS2 ($M_{\rm min}$ is given in unit of
$h^{-1}M_\odot$), respectively, while the value of $\sigma_{\log_{10}M}$ is fixed to $0.35$ for all of the samples. These choices are made such that the resultant clustering signal of the mock galaxies, especially the amplitude of the power spectra at small $k$ becomes roughly consistent with the observation (see the next subsection for more detail). We assume that the populated mock galaxies are located at the center-of-mass position of the core particles determined by \textsc{Rockstar}. Similarly, we assign the the center-of-mass velocities of the same core particles to the mock galaxies, which are used when we displace the positions of mock galaxies to redshift space \citep{2020PhRvD.101b3510K}.
\begin{figure}
\begin{center}
 \includegraphics[width=0.48\textwidth]{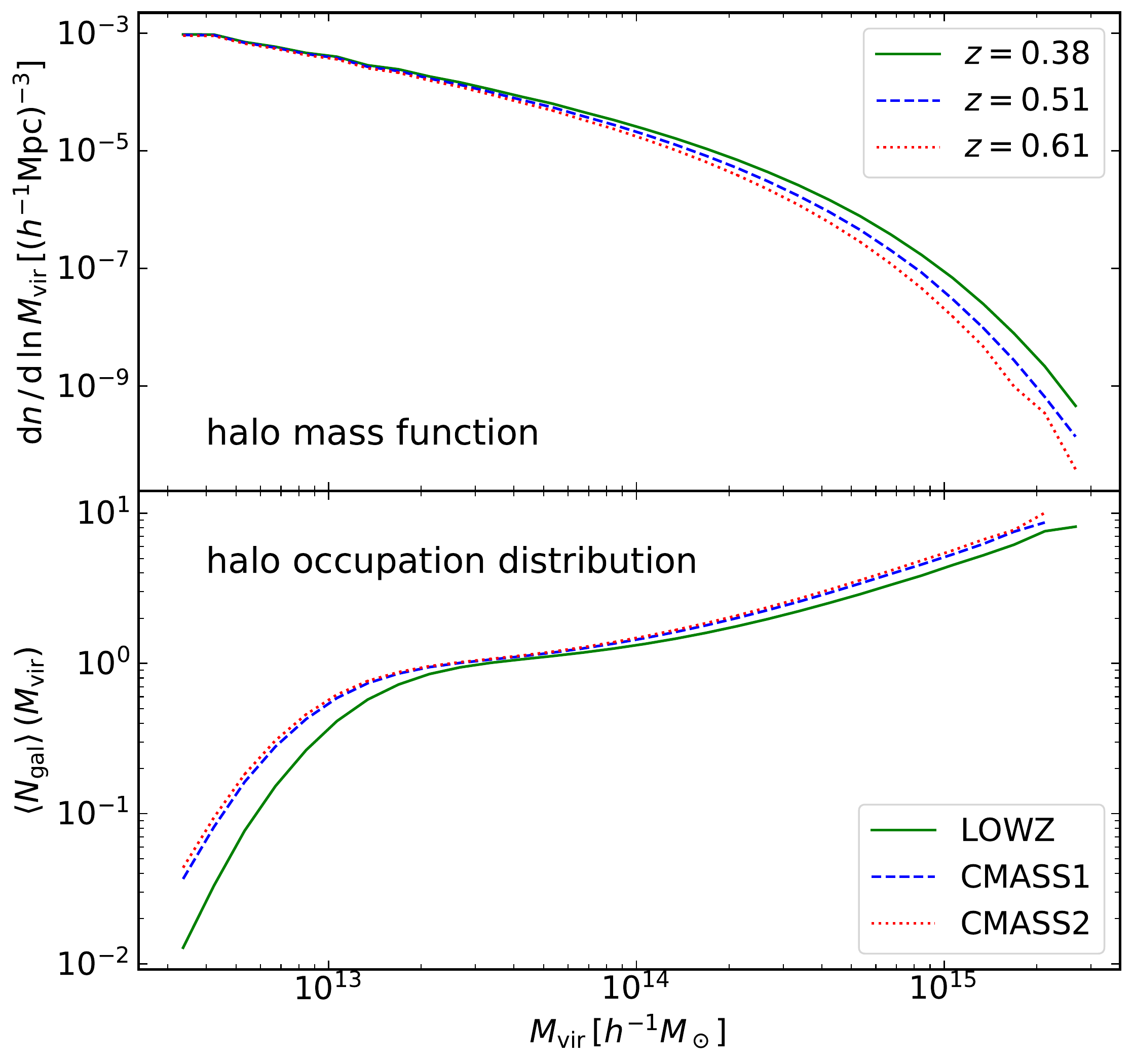}
 \end{center}
\caption{The abundance of halos per unit logarithmic mass interval (upper) and the mean number of mock galaxies per halo (lower) as a function of the virial mass of halos. The mean of the ten random realizations are shown at three output redshifts of the simulations as indicated by the figure legend.}
\label{fig:HOD}
\end{figure}

We show the abundance of (central) halos as well as the mean number of galaxies per central halo as a function of the virial mass in Fig.~\ref{fig:HOD}. Here, we
define ``central'' halos from the \textsc{Rockstar} catalog as those satisfying the condition that
any other halo is not more massive than the halo of interest to within a sphere of radius
$R_{\rm vir}^{\rm cen}$, where $R_{\rm vir}^{\rm cen}$ is the virial radius of the central halo. Note that an isolated halo is also identified as a central halo according to this definition. On the other hand, the halos which reside around a more massive neighbor to within the neighbor's viral radius are identified as ``satellite'' (sub)halos.
The
particular definition does not really affect our mock galaxy catalog due to our recipe (Eq.~\ref{eq:HOD}) to populate galaxies.
The lower panel of Fig.~\ref{fig:HOD} shows the average number of mock galaxies in central halos, i.e. the halo occupation distribution (HOD),
as a function of central halo mass.
Note that unlike the standard
HOD
prescription,
the HOD of our mock catalog
is not given a priori,
and rather is
\textit{measured} from the mocks with the central/satellite split.
Nevertheless, the shape of HOD in our mock catalogs
looks similar to what can be found in the literature, e.g., \citep{2011ApJ...728..126W,2015ApJ...806....2M}.
There appear two regimes; halos around
the soft cutoff near $M_{\rm vir}=10^{13}\,h^{-1}M_\odot$
host only one galaxy (i.e., a central galaxy), while massive halos
above $10^{14}\,h^{-1}M_\odot$ receive
a significant contribution from satellite galaxies, displaying a power-law like form in the HOD.

\subsection{Measurement of the mock signal and error}
\label{subsec:measurement}
We here describe the method to measure the power spectra and estimate the data covariance from the mock galaxy catalogs.

The measurement is done based on FFT of the density field. We first assign the mock galaxies in redshift space to $n_\mathrm{g}^3 = 2048^3$ grid points using the Cloud in Cell (CIC) interpolation scheme. We employ the distant observer approximation in the mapping to the redshift space. We follow Ref.~\cite{Sefusatti16} to correct for the aliasing effect \cite{jing05}, by the so-called interlacing method. To do this, we prepare another density grid but with mass assignment done after shifting the galaxy positions by half the grid size along all of the three Cartesian axes and then corrected for the phase shift by multiplying an appropriate factor to the field in Fourier space. By taking the average of the two density grids, the original and the interlaced, we can get rid of the aliasing effect due to the odd images, which would give the dominant aliasing source to standard cosmological power spectra with decaying amplitude toward higher wavenumbers. The effect of the CIC window function will eventually be removed later in Eq.~(\ref{eq:estimator}).

We then reinterpret the wavenumbers by taking account of the AP effect. Namely, we rescale the fundamental modes along the each of the three axes as
\begin{eqnarray}
&&\tilde{k}_{\mathrm{f},x} = \tilde{k}_{\mathrm{f},y} = \frac{D_\mathrm{A}^\mathrm{(true)}(z)}{D_\mathrm{A}^\mathrm{(fid)}(z)}k_f,\nonumber\\
&&\tilde{k}_{\mathrm{f},z} = \frac{H^\mathrm{(fid)}(z)}{H^\mathrm{(true)}(z)}k_f,\label{eq:AP}
\end{eqnarray}
where $k_\mathrm{f} = 2\pi / L$ is the original fundamental mode in the absence of AP effect. In the above, we take the $z$ direction in the simulation box as the line-of-sight direction, and
the upper scripts, (true) and (fid),
indicate that
the comoving angular diameter distance, $D_\mathrm{A}(z)$, or the Hubble expansion rate, $H(z)$, are calculated assuming the correct, blinded cosmological parameters, and a fiducial cosmological parameters, respectively. Here, we adopt a flat $\Lambda$CDM cosmology with $\Omega_\mathrm{m}^\mathrm{(fid)}=0.3$ as the fiducial cosmology, and this information is shared with the two US analysis teams.
$\Omega_\mathrm{m}^\mathrm{(fid)}$
that was used to create the mock catalogs should not be confused
with the true cosmological parameter
$\Omega_\mathrm{m}$, which was
used in the simulations
and which was kept in secret to
the analyzing teams.

The Japan Team then estimates the first three non-zero multipole moments, monopole ($\ell=0$), quadrupole ($\ell=2$) and hexadecapole ($\ell=4$) by taking weighted averages of the squared Fourier modes:
\begin{eqnarray}
\hat{P}_\ell (k_i) = \frac{2\ell+1}{N_i}\sum_{\tilde{\bfk} \in \mathrm{bin}\,i} \mathcal{P}_\ell(\mu_{\tilde{\bfk}})\hat{P}(\tilde{\bfk}),\label{eq:estimator}
\end{eqnarray}
with
\begin{eqnarray}
\hat{P}(\tilde{\bfk}) = \frac{\tilde{V}\left|\delta_{\tilde{\bfk}}\right|^2-\tilde{P}_\mathrm{shot}(\tilde{\bfk})}{W_\mathrm{CIC}^2(\tilde{\bfk})},
\end{eqnarray}
where the distorted volume $\tilde{V}$ is given by
\begin{eqnarray}
\tilde{V} = \left(\frac{D_\mathrm{A}^\mathrm{(fid)}(z)}{D_\mathrm{A}^\mathrm{(true)}(z)}\right)^2\frac{H^\mathrm{(true)}(z)}{H^\mathrm{(fid)}(z)} \, L^3,\label{eq:AP_vol}
\end{eqnarray}
analogously to Eq.~(\ref{eq:AP}) to account for the AP effect, the summation runs
over
wavevectors $\tilde{\bfk}^{\mathrm{T}} = (\tilde{k}_{\mathrm{f},x} i_x, \tilde{k}_{\mathrm{f},y} i_y, \tilde{k}_{\mathrm{f},x} i_z)$ specified by an integer vector $(i_x,i_y,i_z)$, $\mathcal{P}_\ell$ denotes the $\ell$-th order Legendre polynomial, $\mu_{\tilde{\bfk}}$ is the cosine between the wavevector
$\tilde{\bfk}$ and the $z$-direction,
and $N_i$ stands for the number of Fourier modes contained
in the $i$-th wavenumber bin. In the above, we have subtracted the shot noise, $\tilde{P}_\mathrm{shot}$, from the measured power spectrum\footnote{Notice that this contribution is coming from the zero-lag correlator inherent in point processes and thus exactly $1/n_\mathrm{g}$ for any tracers with a number density $n_\mathrm{g}$. However, on the modeling side, the stochastic contribution in galaxy spectra uncorrelated with large scale density fluctuations is sometimes also referred to as the shot noise. In this definition, it is well known that the shot noise, i.e., the level of stochasticity, can deviate from the $1/n_\mathrm{g}$ Poissonian noise. While we omit this in the analyses shown in the main text, the possible impact of treating this as an additional free parameter is discussed in the Appendix}.
We evaluate the shot noise
taking into account the interlacing technique for the aliasing correction and the CIC window function. Denoting
\begin{eqnarray}
\tilde{\kappa}_a = \frac{\pi \tilde{k}_a}{2\tilde{k}_{\mathrm{Ny},a}},
\end{eqnarray}
with $\tilde{k}_{\mathrm{Ny},a} = \tilde{k}_{\mathrm{f},a} n_\mathrm{g}/2$ being the direction-dependent Nyquist frequency ($a=x$, $y$ or $z$), the resultant expression for the wavevector-dependent shot noise contribution is given as
\begin{eqnarray}
\tilde{P}_\mathrm{shot}(\tilde{\bfk}) &=& \sum_{n_x,n_y,n_z: \mathrm{even}} W_\mathrm{CIC}^2(\tilde{\bfk}+2\tilde{\bfk}_\mathrm{Ny}
\bfn^{\mathrm{T}}) \frac{\tilde{V}}{N_\mathrm{gal}},\nonumber\\
&=& \left[\prod_{a=x,y,z} C_a(\tilde{k}_a)\right] \frac{\tilde{V}}{N_\mathrm{gal}},
\end{eqnarray}
with $W_\mathrm{CIC}$ being the CIC window function
\begin{eqnarray}
W_\mathrm{CIC}(\tilde{\bfk}) = \prod_{a=x,y,z} \mathrm{sinc}^{
2}\tilde{\kappa}_a,
\end{eqnarray}
and the final shot-noise correction factor, $C_a$, given as the infinite summation over even integers can be computed analytically as
\begin{eqnarray}
C_a(\tilde{k}_a) = \frac{1}{12}\left(1+\cos\tilde{\kappa}_a\right)^2 \left(2+\cos\tilde{\kappa}_a\right).
\end{eqnarray}
See Ref.~\cite{jing05} for a similar expression but without the interlacing correction that erases the odd images.

The estimator, Eq.~(\ref{eq:estimator}), is computed at 100 wavenumber bins between the first bin edge taken at zero to the final bin edge at $1\,\hMpci$
evenly spaced by
$0.01\,\hMpci$. The representative wavenumber of each bin, $k_i$ in Eq.~(\ref{eq:estimator}),
is
computed as the average of the norm of the wavevectors that actually enter the bin:
\begin{eqnarray}
k_i = \frac{1}{N_i}\sum_{\tilde{\bfk} \in \mathrm{bin}\,i} \left|\tilde{\bfk}\right|.\label{eq:k_i}
\end{eqnarray}
The pairs of numbers, $(k_i,\hat{P}_\ell(k_i))$,
are provided to the analysis teams as the mock measurements, and the above way to compute the representative number
of each $k$ bin is informed to the analysis team.
The data files also contain estimates of the covariance matrix. It is obtained assuming Gaussianity \citep{2020PhRvD.101b3510K}:
\begin{eqnarray}
\mathrm{Cov}_{ij}^{\ell \ell'} &=& \left\langle\left(\hat{P}_\ell(k_i)-\langle\hat{P}_\ell(k_i)\rangle\right)\left(\hat{P}_{\ell'}(k_j)-\langle\hat{P}_{\ell'}(k_j)\rangle\right)\right\rangle,\nonumber\\
&=& \delta_{ij}^\mathrm{K}\frac{(2\ell+1)(2\ell'+1)}{N_i^2}\nonumber\\
&&\times \sum_{\tilde{\bfk} \in \mathrm{bin}\,i}\mathcal{P}_\ell(\mu_{\tilde{\bfk}})\mathcal{P}_{\ell'}(\mu_{\tilde{\bfk}})
\left[P(\tilde{\bfk})+P_\mathrm{shot}\right]^2,\label{eq:covar}
\end{eqnarray}
where $P(\tilde{\bfk})$ is the expectation value of $\hat{P}(\tilde{\bfk})$. The expression reduces to the real-space formula by Ref.~\cite{feldman94} when $\ell=\ell'=0$. In reality, however, we have to make use of a noisy estimate of the power spectrum $\hat{P}(\tilde{\bfk})$ for each wavevector $\tilde{\bfk}$ instead of $P(\tilde{\bfk})$, and this can impact the estimation of the covariance matrix significantly. Therefore, instead of computing Eq.~(\ref{eq:covar}), we first bin the Fourier modes in 10
evenly-spaced
$|\mu_{\bfk}|$ bins and take the average of $\hat{P}(\tilde{\bfk})$ within each bin to suppress the noise. The binned estimates are then used in Eq.~(\ref{eq:covar}), but the summation now runs over bins instead of individual wavevectors, to obtain our estimate of the covariance matrix.

The Japan Team considers two settings for the covariance matrix. The first is to use the volume and the shot noise consistent with the mock simulations. In addition, they provide another estimate scaled to the BOSS DR12 catalogs, by substituting the number density from the observation and then scaling the number of Fourier modes according to the ratio of the surveyed and the simulated volume. The set of estimates, $\hat{P}(k_i)$ and $\mathrm{Cov}_{ij}^{\ell\ell'}$, with the latter now has only diagonal entries with respect to the subscripts, $i$ and $j$, due to the Gaussian approximation, are tabulated for each of the ten random realizations and provided through the website. The Japan Team leaves
the decision to the US Teams on how to exactly use these estimates: which survey specification for the estimation of the covariance matrix to adopt, to combine the ten realization and analyse the averaged spectra just once or to analyse each realization one by one, or to further estimate the non-Gaussian error from the realization-to-realization scatter.

\begin{figure}
\begin{center}
 \includegraphics[width=0.48\textwidth]{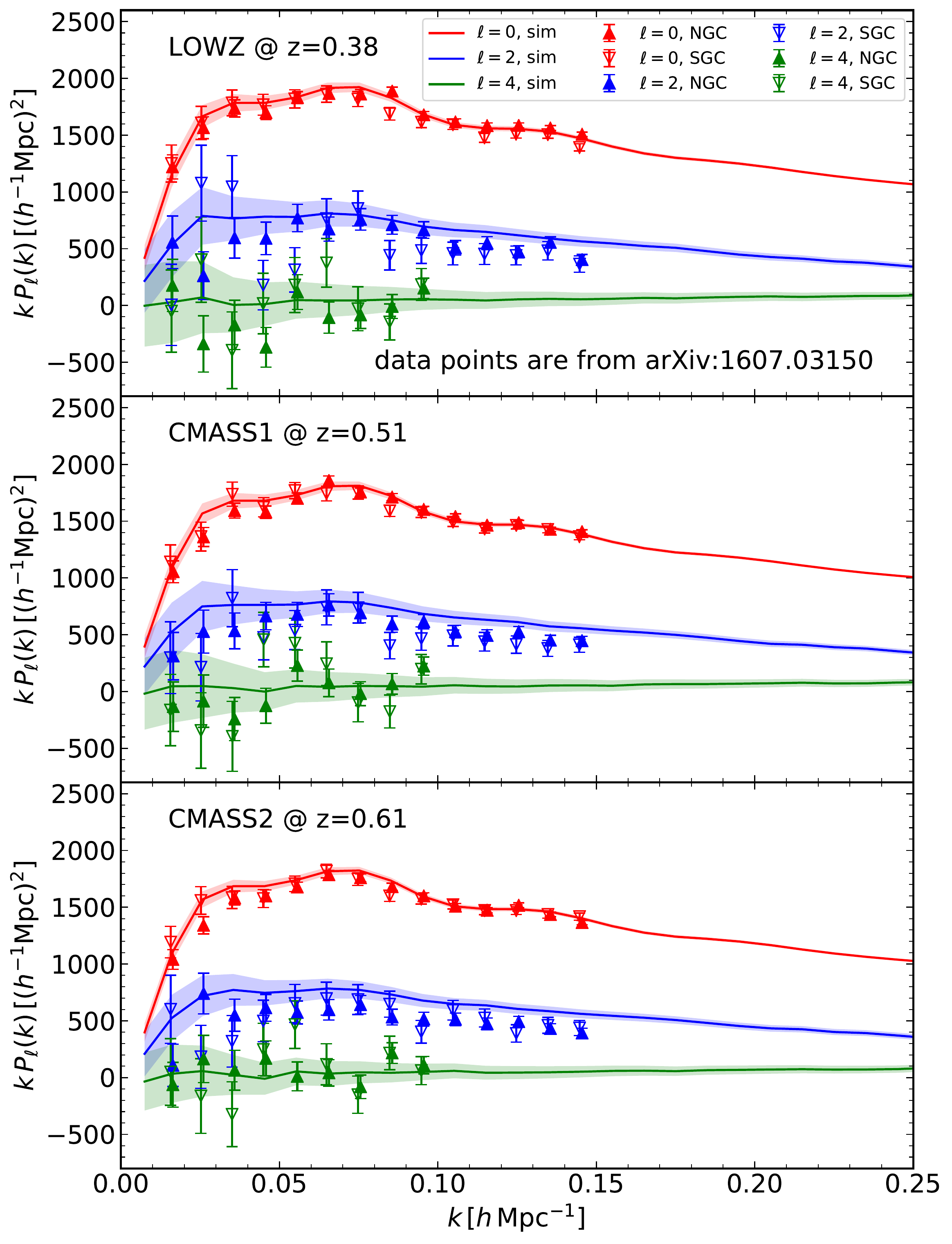}
 \end{center}
\caption{First three multipole moments (monopole, quadrupole and hexadecapole) of the power spectrum in redshift space measured from our mock galaxy catalogs at three redshifts (the solid lines). The $1$-$\sigma$ uncertainty intervals assuming the survey parameter of the SDSS Data Release 12 are shown by the shaded regions. Also shown by the error bars are taken from Ref.~\cite{Beutler17} based on SDSS DR12. For these data points, the measurements from the sample in the North Galactic Cap (NGC) and the South Galactic Cap (SGC) are shown separately by different symbols as indicated by the figure legend. Note that the Alcock-Paczynski effect is artificially induced assuming $\Omega_\mathrm{m}=0.3$ in the redshift-distance conversion. The analysis teams can only access exactly the data vector shown in this figure. The analyses presented in this paper is based on the monopole and the quadrupole moments from the catalog at $z=0.61$.}
\label{fig:multipole}
\end{figure}

We show in Fig.~\ref{fig:multipole} the average multipole moments of the power spectra at the three redshifts corresponding to LOWZ, CMASS1 and CMASS2. The solid lines show the mock measurements, where the shaded region around each line denotes
the $1$-$\sigma$ error scaled to the SDSS BOSS DR12 survey parameters. The three lines in each panel depict the monopole, quadrupole and hexadecapole from top to bottom. Also shown by the symbols with error bars are the actual measurements from the BOSS data by \cite{Beutler17}. The measurements from the North and South Galactic Cap are respectively plotted by the upward and the downward triangles.

Overall, the mock data follows the observed spectra. The monopole moment especially exhibits an excellent agreement, because the model parameters used to distribute the mock galaxies are
chosen to match this moment.
There is, however, small mismatch in the quadrupole moment: the observed data shows a stronger damping behaviour to the higher wavenumbers. It is out of the scope of the current investigation to see if or not this can be alleviated by further tuning the model parameters without spoiling the success in the monopole. This is nontrivial since the cosmological parameters adopted in the mock simulations could be off from the true unknown parameters governing our Universe, or the recipe to populate
mock galaxies might not be flexible enough to meet the reality.

\section{Theoretical template}
\label{sec:theory}

In this section we describe the implementation of the theoretical model by the
two teams participating in the cosmological
analysis challenge. The employed methodologies are almost
identical to the ones used in the analysis of the
actual BOSS data by the same teams
\citep{DAmico:2019fhj,Ivanov:2019pdj,Colas:2019ret}.

Both teams participating in the PT challenge use, essentially, the same theoretical template.
However, there are differences in the implementation of IR resummation, the choice of nuisance  parameters and their priors.
Besides, the two teams use absolutely independent pipelines
based on different software.
This section describes in detail the pipelines used by
the two teams and focuses on methodological differences.

\subsection{Common basis for the EFT formulation}
\label{subsec:common}

On general grounds, it is believed that any physical system has a {\it unique} and {\it correct} description at long wavelengths where the microscopical details of the physical system under consideration can be encoded in just a few coefficients of the terms in the equations of motion. In the context of the long-distance universe, this description is believed to be the Effective Field Theory of Large-Scale Structure (EFTofLSS)~\cite{Baumann:2010tm,Carrasco:2012cv}. The originality of the EFTofLSS with respect to other pre-existing perturbative methods that were applied in the context of LSS is two-fold. First is the presence of suitable terms in the equations of motion that encode the effect of short-distance non-linearities and galaxies at long distances, and that cannot be predicted without detailed knowledge of galaxy physics, and therefore are generically fit to observations. Second, the equations of motion in the EFTofLSS have non-linear terms that are proportional to some parameters. Due to the many phenomena that control the evolution of our universe, there are several of these parameters, such as the size of the density perturbation or the ratio of a given wavelength with respect to the size of the displacements induced by short distance modes~\cite{Senatore:2014via}. For all of these parameters but one, an iterative solution is performed. Instead for one parameter, the one encoding the effect of long wavelength displacements, a non-linear solution is performed, which goes under the name of IR-Resummation~\cite{Senatore:2014via,Baldauf:2015xfa,Senatore:2017pbn,Lewandowski:2018ywf,Blas:2016sfa}. Different incarnations of the EFTofLSS make this expansion more or less manifest. For example, the Lagrangian-space EFTofLSS~\cite{Porto:2013qua} automatically solves non-linearly in the effect of long-displacements, and so, it is identical to the Eulerian EFTofLSS that we use here after this has been IR-Resummed~\cite{Senatore:2014via}.

In the EFTofLSS, the description of the clustering of galaxies in redshift space is performed in the following way. First, the dark matter and baryonic fields are described in terms of fluids with a non-trivial stress tensor. Galaxies are biased tracers, in the sense that, if $\delta_g$ is the galaxy overdensity, we have that~\cite{Senatore:2014eva}
\begin{align}\label{eq:biasgeneral}
&\delta_g(x,t)=\sum_n\int dt' K_n(t,t')\, \tilde {\cal{O}}_n(x_{\rm fl}, t')\\ \nonumber
&\qquad\quad=\sum_{n,m} b_{n,m}(t)\, {\cal{O}}_{n,m}(x, t)
\end{align}
where $\tilde{\cal{O}}_n$ are all possible fields, such as, for example, the dark matter density, that, by general relativity, can affect the formation of galaxies.  $K_n(t,t')$ are some kernels that relate how a field at a certain time affects the galaxies at later times, and $x_{\rm fl}$ is the location at time $t'$ of the fluid element that is at $x$ at time $t$. The last step of the above equation can be performed using the perturbative expression for the matter and baryonic fields. In fact, in perturbation theory the time- and space-dependence parts factorize in a form, schematically, given by $\delta(\vec k,t)\sim \sum_n f_n(t) \delta^{(n)}(\vec k)$, where $\delta^{(n)}$ is order $n$ in the expansion parameters. This allows us to define the biases $b$ as $b_{n,m}(t)\sim \int dt' K_n(t,t') f_m(t')$. This provides the first complete parametrization of the bias expansion, though many earlier attempts were made and substantial but partial successes were obtained.

Next, we need to describe the observed density field in redshift space. This is a combination of the density field in configuration space and
density times powers of the velocity field of galaxies, such as $\rho(\vec x,t) v(\vec x,t)^i, \rho(\vec x,t)v^i(\vec x,t) v_i(\vec x,t),\ldots$. Again, these short-distance-dependent terms are described as above as biased tracers of the density and baryonic fields~\cite{Senatore:2014vja}.

Because of what we just discussed,
the range over which different implementations of the EFTofLSS can differ is extremely limited:
they may choose a different basis for the EFT-parameters, they may add an incomplete, and therefore different, set of higher-order conterterms to partially include the effect of some higher order calculation that was not performed, or they may have different implementations or approximations for the IR-Resummation. We are going to list them in detail next.

\subsection{Group dependent implementation}
\label{subsec:group}

Although both teams use the same theoretical model, there are several
important
methodological differences. Moreover, the two groups have made very different choices in the model implementation and numerical algorithms. This section describes in detail the pipelines used by the two teams.

\subsubsection{East Coast Team}
\label{subsubsec:TeamEC}

The East Coast Team used only the monopole and the quadrupole in the analysis.
The East Coast Team analyzed the challenge data with and without the hexadecapole moment and found identical constraints.\footnote{On the scales of interest the hexadecapole signal is dominated by leakage contributions from the monopole
and quadrupole. These contributions appear due to discreteness effects, i.e.~because the monopole and quadrupole are not exactly orthogonal to the hexadecapole on a finite grid.
Even with the gigantic volume of the challenge simulation and
the wide binning
the hexadecapole moment happened to be dominated by the systematic leakage from lower multipole moments.
}
Given these reasons, the East Coast Team refrained from using the hexadecapole moment in the baseline analysis.

The theoretical model used by the East Coast Team for these two multipoles can be written schematically as
\be
P_\ell(k) = P_\ell^{\rm tree}(k) + P_\ell ^{\rm loop} (k) + P_\ell^{\rm ctr}(k) +
P^{\nabla^4_z \delta}_\ell(k)
\;.
\ee
The tree-level contribution is given by the Kaiser formula \cite{kaiser87}. The loop corrections are calculated using the standard one-loop power spectra for dark matter and biased tracers (see e.g., \cite{Bernardeau:2001qr,Blas:2015qsi,Desjacques18} and references therein). The bias model consists of the following bias operators \cite{Assassi:2014fva,Mirbabayi:2014zca,Senatore:2014eva}
\be
\delta_g (\k)= b_1 \delta (\k) + \frac {b_2}2 \delta^2 (\k) + b_{\mathcal G_2} \mathcal G_2(\k) \;,
\ee
where the momentum-space representation of $\mathcal G_2$ operator is given by
\be
\mathcal G_2(\k) = \int \frac {d^3\p}{(2\pi)^3} \left[ \frac{(\p\cdot(\k-\p))^2}{p^2|\k-\p|^2} - 1 \right] \delta(\p) \delta(\k-\p) \;.
\ee
The one-loop power spectrum has one extra bias operator
 multiplied by an additional  parameter $b_{\Gamma_3}$.
 However, this contribution is almost fully degenerate with the counterterms and $\mathcal{G}_2$ operator on the scales of interest. Given this strong degeneracy, the East Coast Team
has set $b_{\Gamma_3}=0$ in the baseline analysis. Running the MCMC chains with and without $b_{\Gamma_3}$, it was checked that this choice does not affect constraints on cosmological parameters.

The standard one-loop counterterms for the monopole and the quadrupole are \cite{Senatore:2014vja}
\be
P_0^{\rm ctr}(k) = -2c^2_0 k^2 P_{11}(k) \;,\quad P_2^{\rm ctr}(k) =-\frac{4f}{3}c^2_2 k^2
P_{11}(k)\,,
\ee
where $f=\mathrm{d}\ln D_+/\mathrm{d} \ln a$ is the logarithmic growth rate, $D_+$ denotes the linear growth factor and $P_{11}(k)$ is the linear power spectrum.
The purpose of these counterterms is to fix the UV-dependence of the loops and to partly take into account the effects of
the fingers-of-God \cite{jackson72}.
The East Coast Team also added an extra $k^4$ term shared between the multipoles,
\be
\label{eq:k4}
P^{\nabla^4_z \delta}(k,\mu) =  - c (\mu k f)^4 (b_1+f\mu)^2 P_{11}(k) \;.
\ee
 This new counterterm takes into account next-to-leading order of the fingers-of-God. Note that on general grounds one also expects the presence of the stochastic contribution of the form \cite{Senatore:2014vja,Perko:2016puo},
\be
P_{\text{RSD, stoch}}= -c_{\epsilon }k^2 \mu^2\,.
\ee
This contribution happens to be very degenerate with the counterterm
\eqref{eq:k4} on the scales of interest for the analysis and it was not included in the model by the East Coast Team.

The East Coast Team has implemented  IR-Resummation and the Alcock-Paczynski effect as explained in detail
in Refs.~\cite{2019JCAP...11..034C,2019arXiv190905277I}.
Importantly, the East Coast
team has used
 the IR resummation algorithm based on the wiggly-smooth decomposition directly in Fourier space \cite{Baldauf:2015xfa,Blas:2016sfa,Ivanov:2018gjr},
which allowed for a significant boost of computational speed.
This scheme
is efficient and numerically stable.
Moreover, it is based on solid
systematic parametric
expansion that guarantees
that the error is under control
at every order of IR resummation.
It was explicitly checked that the residuals introduced by
our procedure are much smaller
than the 2-loop contributions which are not included in the model, in full agreement with theoretical
expectations \cite{Blas:2016sfa,Ivanov:2018gjr}.
The labels that indicate  IR-Resummation and the AP effect were omitted in all equations in this section to avoid clutter. However, the reader should keep in mind that they are always included in the model.

The total number of nuisance parameters used in the blinded analysis of the East Coast Team is 6: three counterterms ($c^2_0$, $c^2_2$, $c$) and three bias parameters ($b_1$, $b_2$, $b_{\mathcal G_2}$).
Since the shot noise contribution has been subtracted from the measured spectra, the corresponding parameter was not fitted, in contrast to Ref.~\cite{2019arXiv190905277I}.
As far as the cosmological parameters are concerned, the basis that was used consists of the dimensionless Hubble constant $h$ ($H_0=h\cdot 100$ km/s/Mpc),
the physical matter density $\omega_\mathrm{m}$,
and the normalization $A^{1/2}$ defined with respect to the best-fit Planck value for the base $\Lambda$CDM cosmology,
\be
\begin{split}
&A^{1/2}\equiv \left(\frac{A_{\rm s}}{A_{{\rm s},\,\text{Planck}}}\right)^{1/2}\,,\\
&
\text{where }
\quad
A_{{\rm s},\,\text{Planck}} = 2.0989\cdot 10^{-9}\,.
\end{split}
\ee
All varied cosmological and nuisance parameters were assigned flat priors without boundaries, i.e. $(-\infty,\infty)$.

The evaluation of perturbation
theory integrals was performed using
the FFTLog method of \cite{Simonovic:2017mhp}
implemented as a module in the \textsc{CLASS} Boltzmann solver \cite{class2,Chudaykin:2020aoj}.
Using the IR-Resummation based on wiggly-smooth decomposition, a single evaluation of a
theory model is of the order $\mathcal O(1)$ sec for high precision settings.
This allows for a new evaluation of the non-linear power spectra at every step of the MCMC chain, which is what is done in the East Coast Team analysis.
The MCMC analysis was performed using the \textsc{Montepython~v3.0} \cite{Audren:2012wb,Brinckmann:2018cvx}
sampler interfaced with the modified version of the \textsc{CLASS} code.
The nuisance parameters were sampled in the ``fast mode'' \cite{Lewis:2002ah} at a negligible computational cost.

Since the $k$-binning
of the challenge spectra is very wide
($\Delta k = 0.01~\hMpci$)
compared to the fundamental mode of the box, the theoretical
predictions had to be properly averaged over each bin. The boundaries of the bins were
estimated using the simulation volume, known to both teams.
The East Coast Team checked that the estimated boundaries allow one to accurately
reproduce the provided weighted means of the $k$-bins and found that averaging the theory over the bin versus evaluating it in the mean can induce roughly $\mathcal O(0.5)\sigma$ shifts in cosmological parameters.

\subsubsection{West Coast Team}
\label{subsubsec:TeamWC}
The implementation of the West Coast Team is the result of a long journey where each of ingredients of the EFTofLSS that is necessary to apply it to data was one-by-one subsequently developed, tested on simulations, shown to be successful. Though not all those results are directly used in the analysis, the West Coast Team, and probably nobody, would simply have never applied the model to the data without those intermediate successes. We therefore find it nice to add, in each instance where the EFTofLSS is applied to data, the following footnote where we acknowledge at least a fraction of those important developments\footnote{The initial formulation of the EFTofLSS was performed in Eulerian space in~\cite{Baumann:2010tm,Carrasco:2012cv}, and then extended to Lagrangian space in~\cite{Porto:2013qua}. The dark matter power spectrum has been computed at one-, two- and three-loop orders in~\cite{Carrasco:2012cv, Carrasco:2013sva, Carrasco:2013mua, Carroll:2013oxa, Senatore:2014via, Baldauf:2015zga, Foreman:2015lca, Baldauf:2015aha, Cataneo:2016suz, Lewandowski:2017kes,Konstandin:2019bay}. Some additional theoretical developments of the EFTofLSS that accompanied these calculations were a careful understanding of renormalization~\cite{Carrasco:2012cv,Pajer:2013jj,Abolhasani:2015mra} (including rather-subtle aspects such as lattice-running~\cite{Carrasco:2012cv} and a better understanding of the velocity field~\cite{Carrasco:2013sva,Mercolli:2013bsa}), of the several ways for extracting the value of the counterterms from simulations~\cite{Carrasco:2012cv,McQuinn:2015tva}, and of the non-locality in time of the EFTofLSS~\cite{Carrasco:2013sva, Carroll:2013oxa,Senatore:2014eva}. These theoretical explorations also include an instructive study in 1+1 dimensions~\cite{McQuinn:2015tva}. In order to correctly describe the Baryon Acoustic Oscillation~(BAO) peak, an IR-resummation of the long displacement fields had to be performed. This has led to the so-called IR-Resummed EFTofLSS~\cite{Senatore:2014via,Baldauf:2015xfa,Senatore:2017pbn,Lewandowski:2018ywf,Blas:2016sfa}.  A method to account for baryonic effects was presented in~\cite{Lewandowski:2014rca}. The dark-matter bispectrum has been computed at one-loop in~\cite{Angulo:2014tfa, Baldauf:2014qfa}, the one-loop trispectrum in~\cite{Bertolini:2016bmt}, and
the displacement field in~\cite{Baldauf:2015tla}. The lensing power spectrum has been computed at two loops in~\cite{Foreman:2015uva}.  Biased tracers, such as halos and galaxies, have been studied in the context of the EFTofLSS in~\cite{ Senatore:2014eva, Mirbabayi:2014zca, Angulo:2015eqa, Fujita:2016dne, Perko:2016puo, Nadler:2017qto} (see also~\cite{McDonald:2009dh}), the halo and matter power spectra and bispectra (including all cross correlations) in~\cite{Senatore:2014eva, Angulo:2015eqa}. Redshift space distortions have been developed in~\cite{Senatore:2014vja, Lewandowski:2015ziq,Perko:2016puo}. Clustering dark energy has been included in the formalism in~\cite{Lewandowski:2016yce,Lewandowski:2017kes,Cusin:2017wjg,Bose:2018orj}, primordial non-Gaussianities in~\cite{Angulo:2015eqa, Assassi:2015jqa, Assassi:2015fma, Bertolini:2015fya, Lewandowski:2015ziq, Bertolini:2016hxg}, and neutrinos in~\cite{Senatore:2017hyk,deBelsunce:2018xtd}. Faster evaluation schemes for evaluation for some of the loop integrals have been developed in~\cite{Simonovic:2017mhp}.}.

 The model for the West Coast Team and the analysis techniques are the same as the one used in~\cite{DAmico:2019fhj,Colas:2019ret}, to which we refer for details. The one-loop redshift-space galaxy power spectrum reads:
\begin{align}\label{eq:powerspectrum}\nonumber
P_{g}(k,\mu) & =  Z_1(\mu)^2 P_{11}(k)  \\ \nonumber
& + 2 \int \frac{d^3q}{(2\pi)^3}\; Z_2(\q,\k-\q,\mu)^2 P_{11}(|\k-\q|)P_{11}(q) \\ \nonumber
&+ 6 Z_1(\mu) P_{11}(k) \int\, \frac{d^3 q}{(2\pi)^3}\; Z_3(\q,-\q,\k,\mu) P_{11}(q)\nonumber \\\nonumber
& + 2 Z_1(\mu) P_{11}(k)\left( c_\text{ct}\frac{k^2}{{ k^2_\textsc{m}}} + c_{r,1}\mu^2 \frac{k^2}{k^2_\textsc{m}} + c_{r,2}\mu^4 \frac{k^2}{k^2_\textsc{m}} \right)\\
&+ \frac{1}{\bar{n}_g}\left( c_{\epsilon,1}+ c_{\epsilon, 2}\frac{k^2}{k_\textsc{m}^2} + c_{\epsilon,3} f\mu^2 \frac{k^2}{k_\textsc{m}^2} \right).
\end{align}
$k^{-1}_\textsc{m}$ controls the bias derivative expansion and we set it to be $\simeq k^{-1}_\textsc{nl}$, which is the scale controlling the expansion of the dark matter  derivative expansion. We set $k_\textsc{nl}=0.7 h {\rm Mpc}^{-1}$. $\bar{n}_g$ is the mean galaxy density.

In the next to the last line of Eq.~(\ref{eq:powerspectrum}), the term in $c_{\rm ct}$ represents a linear combination of a higher derivative bias~\cite{Senatore:2014eva} that appears in Eq.~(\ref{eq:biasgeneral}) and the speed of sound of dark matter~\cite{Baumann:2010tm,Carrasco:2012cv}: $\delta(\vec k,t)\supset k^2 \delta_{\rm lin}(\vec k, t)$. The terms in $c_{r,1}$ and $c_{r,2}$ represent the redshift-space counterterms~\cite{Senatore:2014vja}: $\delta_{\rm redshift}(\vec k,t)\supset k^2 \mu^2 \delta(k,t),\ k^2 \mu^4 \delta(k,t) $. In the last line of Eq.~(\ref{eq:powerspectrum}), we have the stochastic counterterms: $c_{\epsilon,1}$ and $c_{\epsilon,2}$ originate from Taylor
expansion of
Eq.~(\ref{eq:biasgeneral})~\cite{Senatore:2014eva}, while $c_{\epsilon,3}$ originates from the redshift-space expressions~\cite{Senatore:2014vja}.

The redshift-space galaxy density kernels $Z_1,Z_2$ and $Z_3$ are given in Appendix~\ref{sec:galaxykernels}. These kernels depend on the bias coefficients that we define as explained below Eq.~(\ref{eq:biasgeneral}).  By choosing only the linearly-independent ones, this gives rise to the so-called base of descendants. While up to cubic order this base is equivalent to more standard bases, already at quartic perturbative order new terms appear.

The IR-resummation is performed in a numerically efficient way using the original method for configuration and redshift space developed in~\cite{Senatore:2014via,Senatore:2017pbn,Lewandowski:2018ywf}, where all the errors are parametrically controlled by the perturbative order of the calculation ({\it i.e.} no uncontrolled approximations are present)~\footnote{Especially within the observational community, a non-linear treatment of the BAO based on the decomposition of the wiggle and smooth part of the power spectrum has been popular for a long time (see for example~\cite{Eisenstein:2006nj}). However, this Team does not find this decomposition to be under parametric control (i.e. there is no small parameter controlling its correctness). It is possible to go from the original IR-Resummation to the simplified ones based on the decomposition by performing a series of approximations (see Appendix of~\cite{Lewandowski:2018ywf}). Of course, this does not mean that the errors which are introduced are large or significant, as can be {a-posteriori} checked on numerical simulations.}.

We define the following combination of parameters: $c_2 = (b_2 + b_4) / \sqrt{2}$, $c_4 = (b_2 - b_4) / \sqrt{2}$, $c_{\epsilon,\rm mono} = c_{\epsilon,\rm 2} + f c_{\epsilon,\rm 3} / 3$ and $c_{\epsilon,\rm quad} = 2 f c_{\epsilon,\rm 3} / 3$.
As we analyze only the monopole and the quadrupole, we set $c_{r,2} = 0$ since the two redshift-space counterterms are degenerate in this case, but we allow a larger prior on $c_{r,1}$ to absorb the contribution of $c_{r,2}$ in the quadrupole.
Additionally, since the shot noise is known and has been subtracted from the data, we set $c_{\epsilon,1}=0$. This leaves us with the set ($b_1$, $c_2$, $b_3$, $c_4$, $c_{ct}$, $c_{r,1}$, $c_{\epsilon, \rm mono}$, $c_{\epsilon, \rm quad}$) of 8 parameters.
The PT challenge data are precise enough to determine all EFT parameters with no priors. However, we impose the following priors motivated by the fact that all EFT parameters are expected to be $\mathcal{O}(1)$~\footnote{
Notice that the consistency of the EFTofLSS is based on a power counting argument that assumes that the subsequent terms of the perturbative expansion are much smaller than the ones that are kept. In order for this to be the case, it is essential that the physical nuisance parameter are kept $\mathcal{O}(1)$, once the relevant physicals scales have been factorized.
}
:
\begin{equation}
\begin{split}
    & b_1\in [0, 4]_{\rm flat} \, , \quad
    c_2 \in [-4, 4]_{\rm flat} \, , \quad
    b_3 \in 10_{\rm gauss} \, , \\
    & c_4 \in 2_{\rm gauss}\, , \quad
    c_{\rm ct} \in 4_{\rm gauss} \, , \quad
    c_{r,1} \in 8_{\rm gauss}\, , \\
    & c_{\epsilon,\rm mono} \in  2_{\rm gauss}\, , \quad
    c_{\epsilon,\rm quad} \in  4_{\rm gauss}\, .
\end{split}
\end{equation}

As it is evident from Eqs.~(\ref{eq:powerspectrum}) and (\ref{eq:redshift_kernels}), some EFT-parameters appear linearly in the model power spectrum, and therefore appear quadratically in the Likelihood. If we are not interested in the actual value of these parameters, as it is our case, we can marginalize over these parameters analytically, obtaining a marginalized likelihood that is a function of only 3 parameters: $b_1$, $c_2$ and~$c_4$.

Given that the $k$-bins ($\Delta k=0.01h/{\rm Mpc}$) contain many fundamental modes, the West Coast Team averages the predictions of the model over each bin. As a check, the Team verified that the provided effective $k$ of the bin was correctly reproduced.

In terms of the cosmological parameters, the West Coast Team has parameterized their analysis in terms of the dimensionless Hubble constant $h$ ($H_0=h\cdot 100$ km/s/Mpc), the present-day matter density fraction $\Omega_\mathrm{m}$, and the normalization of the power spectrum $A_\mathrm{s}$.
The evaluation of the perturbation theory integrals were performed either by direct numerical integration, or by the FFTLog method of~\cite{Simonovic:2017mhp}, obtaining the same result.

\section{Results of blinded analysis}
\label{sec:res_blind}

In this section we display the results obtained by the two teams.
The input values of the cosmological parameters were unblinded
after each team has submitted its results
for consensus data cuts. We present these results in the original
form prepared by either team independently.
Both teams have chosen to analyze
the mean power spectrum (at $z=0.61$)
over 10 realizations
with the covariance estimated from the inverse sum of covariances for
10 single boxes,
\be
\bar C = \left(\sum_i C_i^{-1}\right)^{-1}\,,
\quad \bar P =
\bar C * \sum_i C^{-1}_i P_i\,,
\ee
where $P_i$, $C_i$ are the power spectrum and covariance of
the $i$'th box and ${\bar P}$, ${\bar C}$ are the final mean and covariance that have been analyzed.

This procedure ensures
that the analysis is
approximately equivalent to fitting
the spectrum from
a single simulation box of 566~($h^{-1}\text{Gpc})^3$ volume.
We stress that the obtained statistical errors
on cosmological parameters
correspond to the total volume of 10 simulation
boxes, i.e.~566~($h^{-1}\text{Gpc})^3$.

\subsection{East Coast Team}
\label{subsec:TeamA}

Although the East Coast Team submitted its baseline results for the average over 10 challenge boxes at $z=0.61$, they have also analyzed the data for other redshifts and found consistent results
across all challenge spectra. Prior to unblinding, the East Coast Team has submitted results for 8 different
evenly-spaced
values of $k_{\rm max}$ in the range $(0.08-0.2)\;\hMpci$.

\begin{figure*}[ht]
\begin{center}
\includegraphics[width=0.49\textwidth]{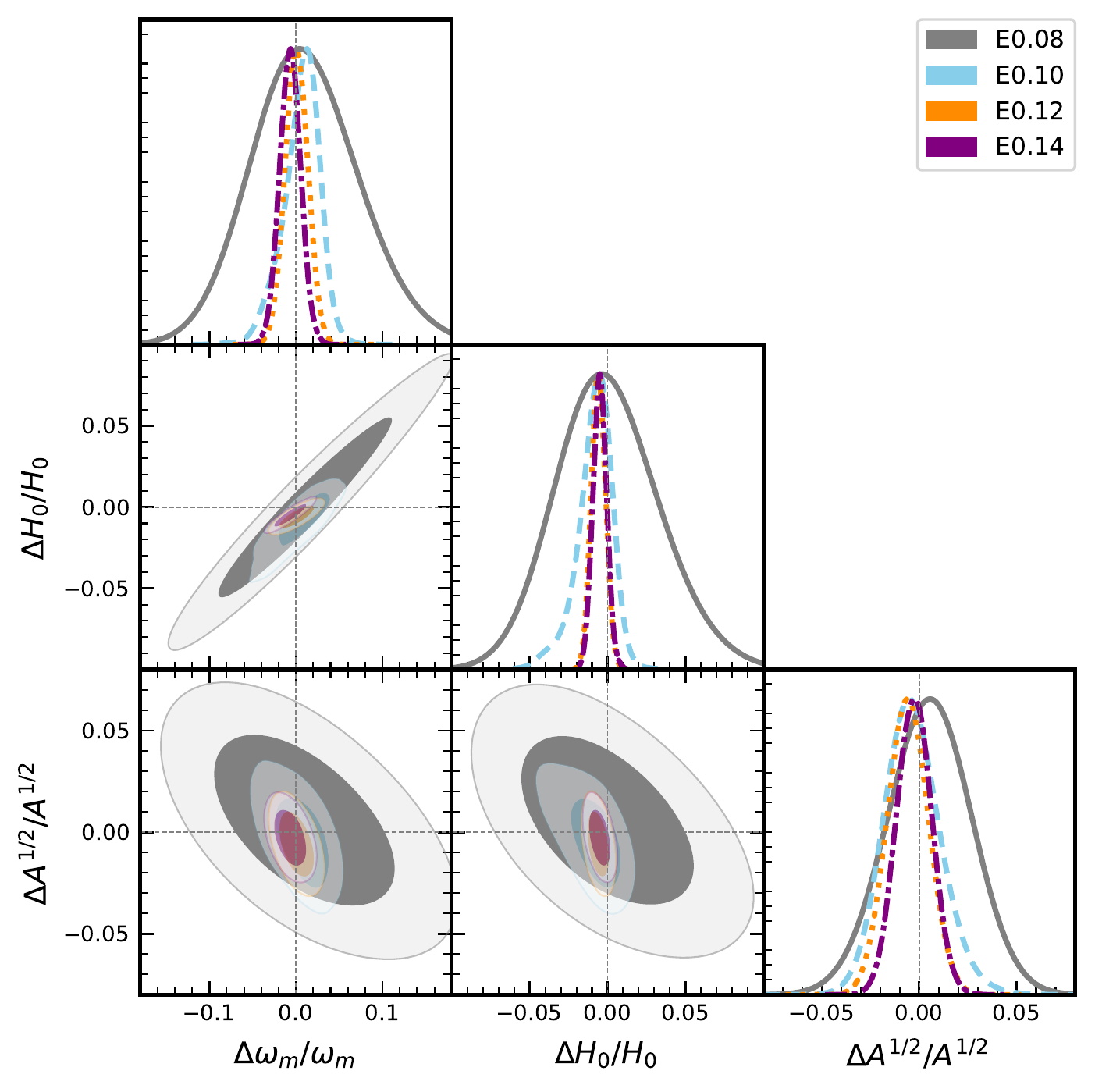}
\includegraphics[width=0.49\textwidth]{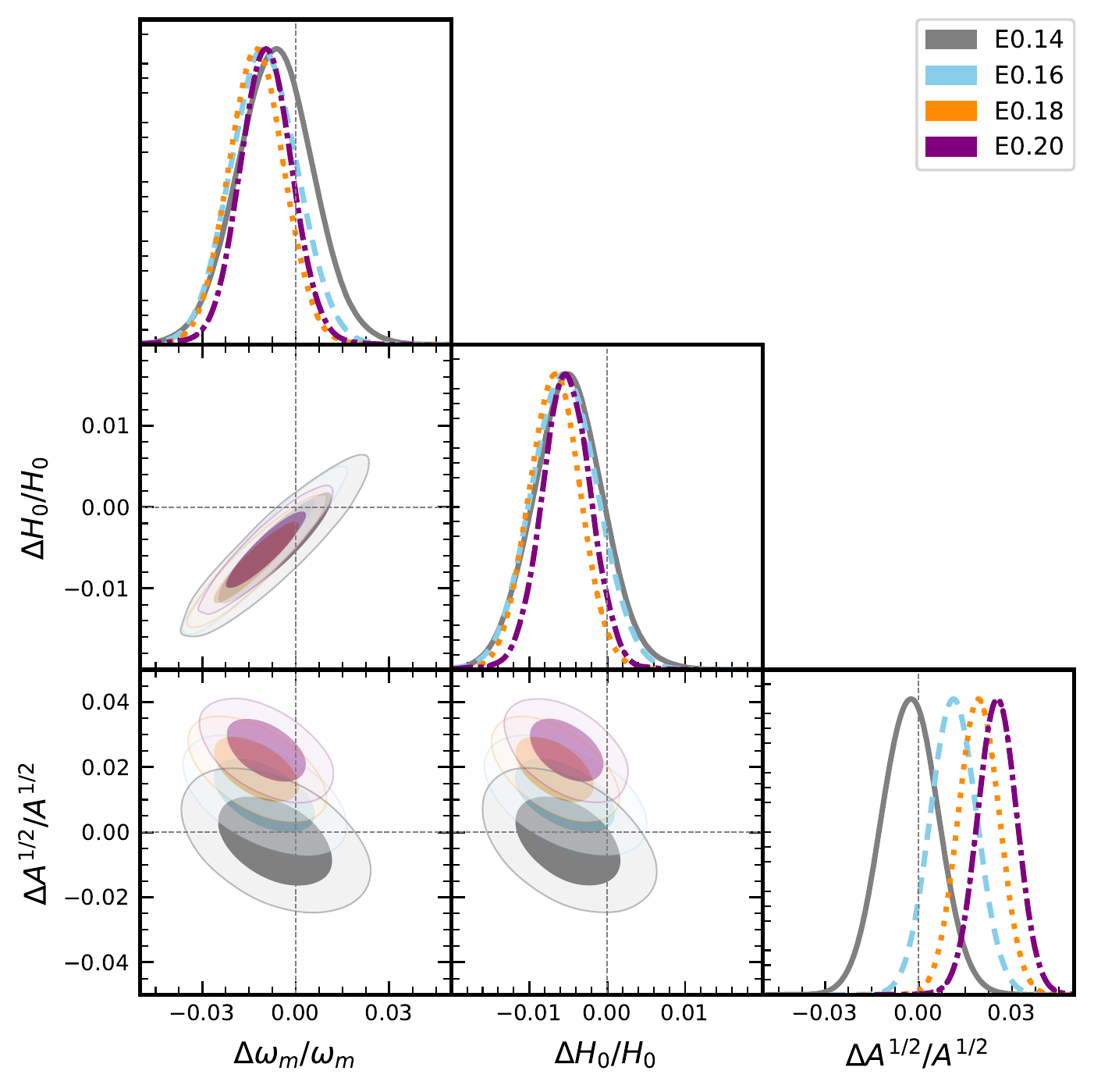}
\end{center}
\caption{Marginalized posteriors for the three varied cosmological parameters as a function of $k_{\rm max}$ (quoted in $\hMpci$ in the figure legend)
obtained by the East Coast Team.
Dashed lines mark the input parameters which were revealed once the Team submitted its final result.}
\label{fig:contours1}
\end{figure*}

\begin{figure*}[ht]
\begin{center}
\includegraphics[width=0.49\textwidth]{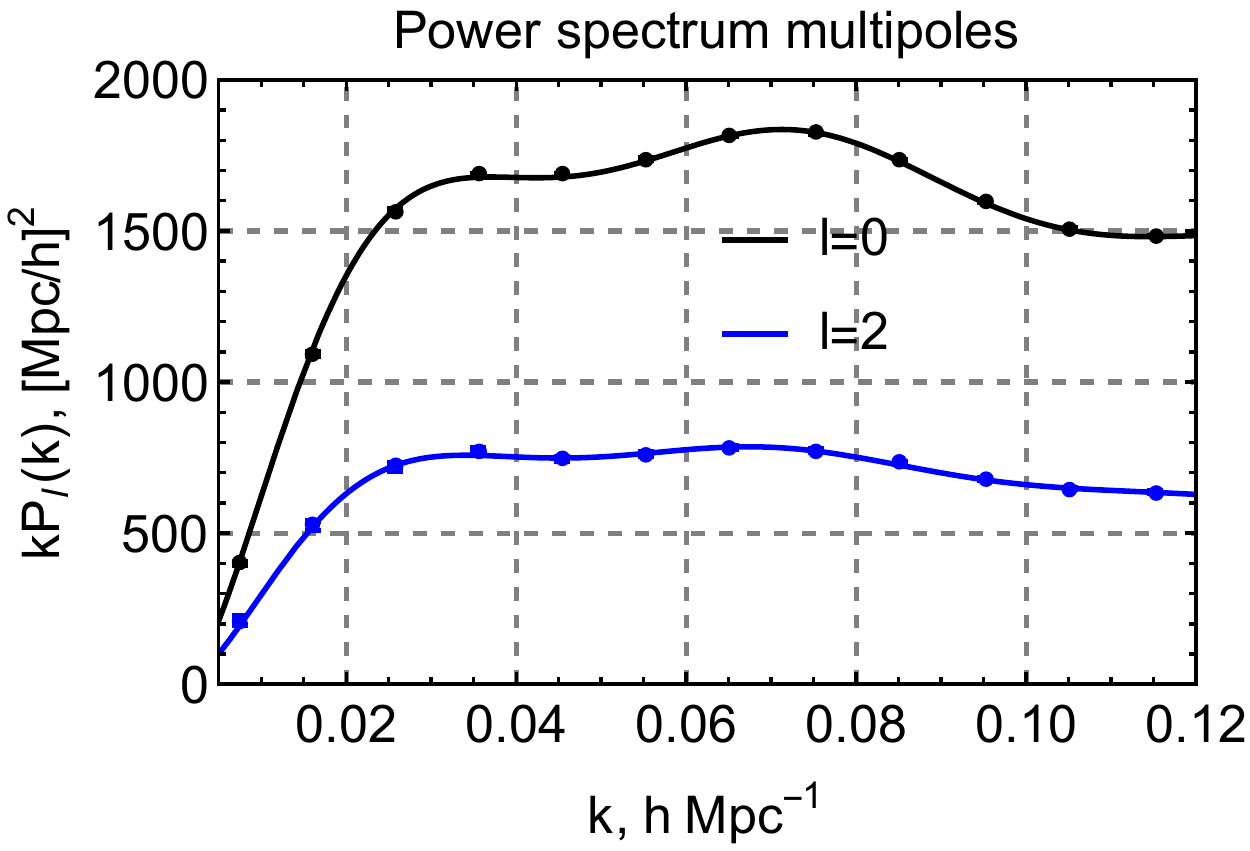}
\centering \includegraphics[width=0.49\textwidth]{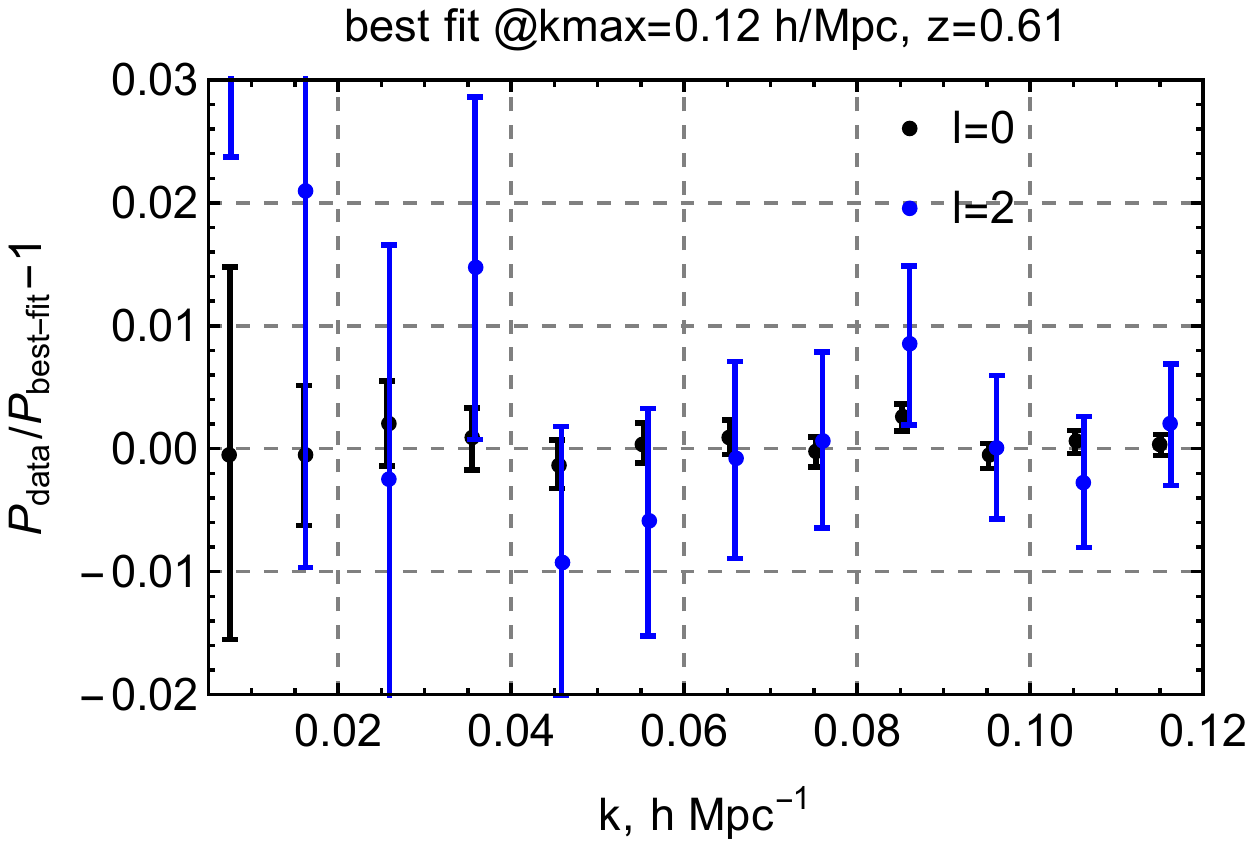}
\centering \includegraphics[width=0.49\textwidth]{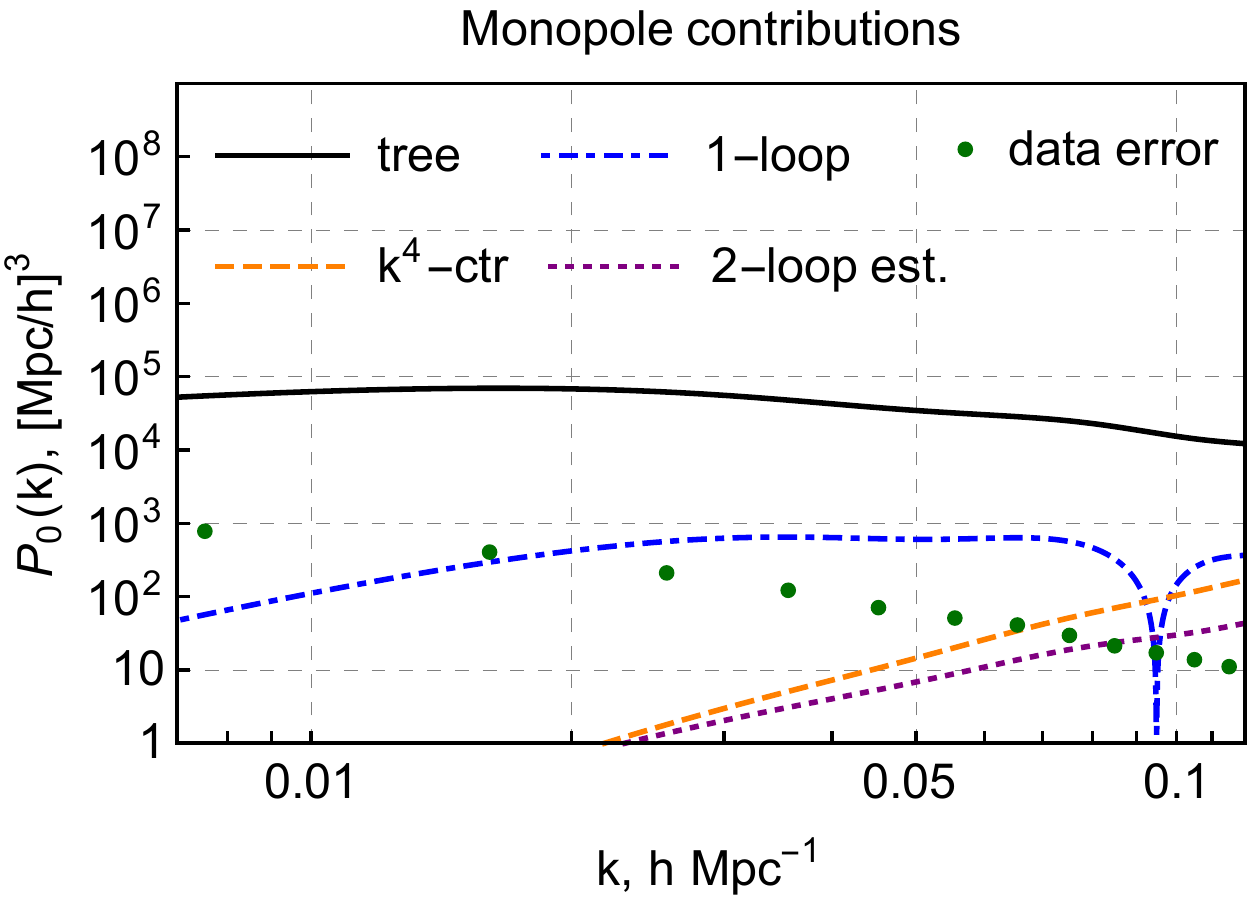}
\centering \includegraphics[width=0.49\textwidth]{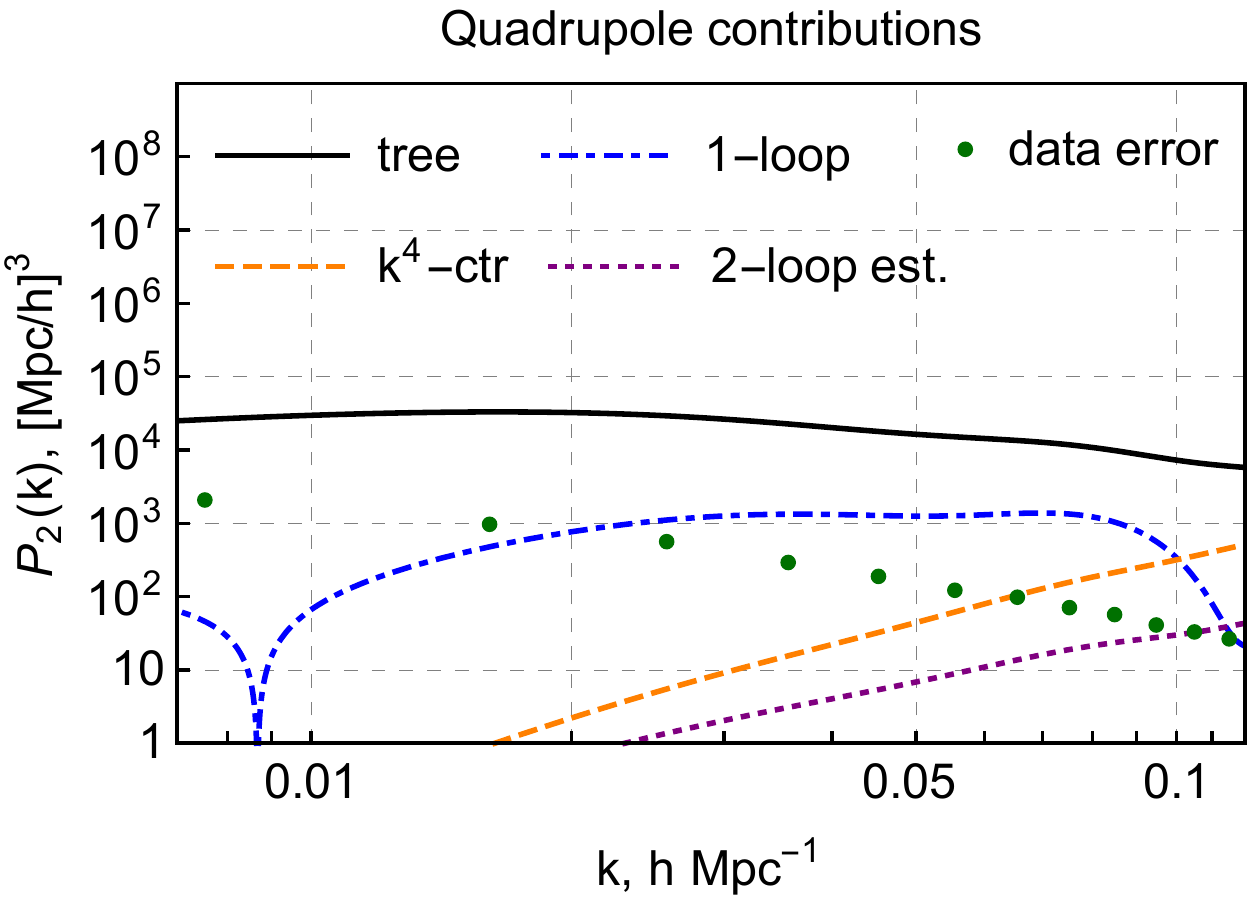}
\end{center}
\caption{
\textit{Upper panel}: Comparison of the data for the monopole and the quadrupole (the error bars are there,
albeit barely visible) with the best-fit model (left panel) obtained by the East Coast Team.
The residuals for the monopole and the quadrupole for the best-fit model with $\chi^2/{\rm dof}= 12/(24-9)$ (right panel).
Note that the quadrupole
data points are slightly shifted for better visibility.
\textit{Lower panel}:
Different contributions to the monopole (left panel) and quadrupole (right panel) power spectra. The data errors and
the two-loop estimate are also displayed.
We plot the absolute values, some terms are negative.}
\label{fig:breakdown1}
\end{figure*}

The marginalized posteriors for
the three cosmological parameters are shown in Fig.~\ref{fig:contours1} for several choices of $k_{\rm max}$.
Between $k_{\rm max}=0.08\;\hMpci$
and $k_{\rm max}=0.14\;\hMpci$
the different contours are compatible within $1\sigma$.
When pushing to higher values of $k_{\rm max}$, the shifts in the central values of the posterior distributions become significant.
Note that for $k_{\rm max}>0.14\;\hMpci$
the contours of $h$ and $\omega_\mathrm{m}$
remain consistent even though the other parameter exhibit clear shifts.
The East Coast Team quoted its final results for a conservative choice of $k_{\rm max}=0.12\;\hMpci$
because this is the scale up to which the Team believed the theoretical modeling
is sufficiently accurate
given sub-percent statistical error bars and the size of neglected nonlinear corrections (see Fig.~\ref{fig:breakdown1}, in which we display an estimate of the two-loop
correction from \cite{Baldauf:2016sjb}).
The 1d marginalized limits for the cosmological and parameters
and the linear bias $b_1$
are given in Table~\ref{tab:res}.
After the true parameters were unblinded, the
values obtained by the East Cost Team were
replaced by relative differences.
For convenience, the values of $\sigma_8$, $\Omega_\mathrm{m}$ and $\ln(10^{10}A_\mathrm{s})$
derived from the East Coast Team MCMC chains are also quoted.
As we have seen after unblinding, the true values of $\omega_\mathrm{m}$ and $h$ reside within $2\sigma$ posterior regions even at
$k_{\rm max}=0.2~\hMpci$,
while the clustering amplitude measurement is consistent up to $k_{\rm max}=0.14~\hMpci$.
Importantly, the Team
has also inferred a correct value of the linear bias\footnote{
The true value of the linear bias was estimated as follows.
The Japan team has measured the real space
matter-matter auto-spectra along with the galaxy-matter
cross spectrum.
Then, we took
the ratio,
$b_1=P_{gm}/P_{mm}$
evaluated in the very first $k$-bin
averaged over the ten realizations as an estimate of the bias parameter.
}
coefficient $b_1$.

\begin{table*}[t!]
\begin{tabular}{|c|c|c|}
 \hline
$  k_{\rm max} =0.12~\hMpci$ & best-fit & mean  $\pm 1\sigma$  \\ [0.5ex]
 \hline\hline
 $ \Delta A^{1/2}/A^{1/2} \cdot 10^{2}$ & $-0.15$ & $-0.16\pm 1.0$  \\
 \hline
 $\Delta h/h \cdot 10^{2}$  & $-0.55$ & $-0.59 \pm 0.46$  \\
 \hline
 $\Delta \omega_\mathrm{m}/\omega_\mathrm{m}\cdot 10^2$  & $0.2 $ & $0.15 \pm 1.4$  \\
 \hline
  $\Delta b_1/b_1\cdot 10^2$  & $0.20$ & $0.22 \pm 1.2$  \\
  \hline \hline
 $\Delta \Omega_\mathrm{m}/\Omega_\mathrm{m} \cdot 10^2$  & $1.3$ & $1.2 \pm 0.9$  \\ \hline
  $\Delta \ln(10^{10}A_\mathrm{s})/\ln(10^{10}A_\mathrm{s})\cdot 10^2$  & $-0.098$ & $-0.11 \pm 0.69$  \\ \hline
  $\Delta \sigma_8/\sigma_8 \cdot 10^2$ & $-0.094$& $-0.022 \pm 0.92$  \\ \hline
\end{tabular}
 \caption{
 The baseline results obtained by the East Coast Team for  $k_{\rm max}=0.12\hMpci$ at $z=0.61$. Only the cosmological parameters and $b_1$ are shown. Note that $\Omega_{\rm m}$, $\ln(10^{10}A_\mathrm{s})$ and $\sigma_8$ in the lower disjoint table shows the results for derived parameters.
 }
 \label{tab:res}
\end{table*}

Fig.~\ref{fig:breakdown1} shows the comparison of the best-fit model at $k_{\rm max}=0.12~\hMpci$
to the data and the residuals.
The quality of the fit is quite good, $\chi^2/{\rm dof}= 12/(24-9)$.
It is consistent with the hypothesis that the data follow the $\chi^2$-distribution with $15$ degrees of freedom.
The lower panel of Fig.~\ref{fig:breakdown1} displays
a breakdown of different contributions to the best-fit model.
The linear theory contribution dominates on all scales, which is consistent with the applicability of perturbation theory.
Towards $k_{\rm max} = 0.12\;\hMpci$
the loop corrections (including the $k^2$-counterterms) become progressively important.
Note that the one loop corrections are detectable already
on very large scales, $\sim 0.02~\hMpci$.
The $k^4$-counterterm is important only for the quadrupole around~$k_{\rm max}= 0.12\;\hMpci$, where it dominates over the other loop corrections.

\subsection{West Coast Team}
\label{subsec:TeamB}

\begin{figure*}[ht]
\begin{center}
\includegraphics[width=0.49\textwidth]{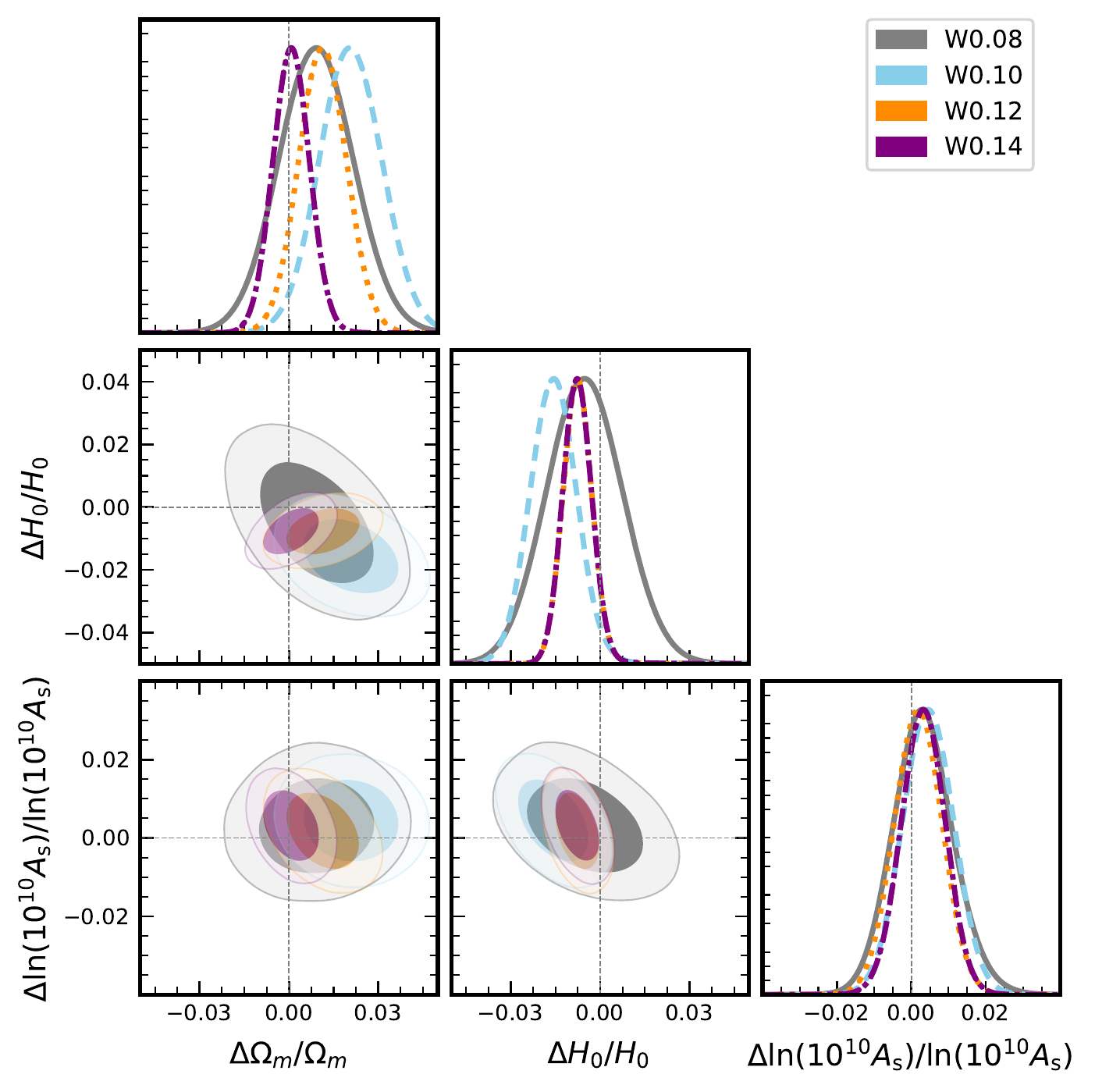}
\includegraphics[width=0.49\textwidth]{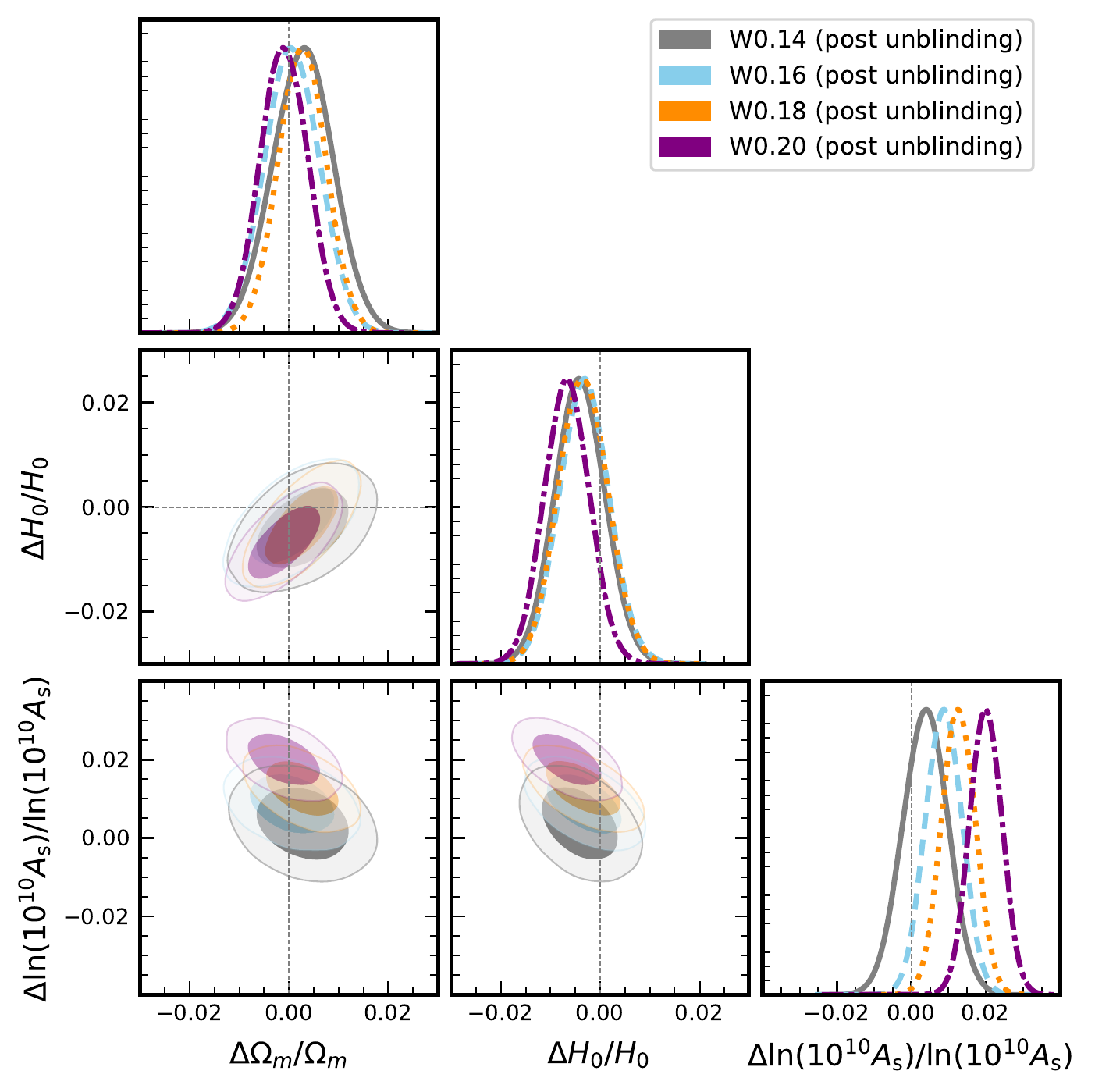}
\end{center}
\caption{Marginalized posteriors for the three varied cosmological parameters as a function of $k_{\rm max}$ (quoted in
$\hMpci$
in the figure legend)
obtained by the West Coast Team.
Dashed lines mark the input parameters which were revealed once the Team submitted its final result,
similarly to Fig.~\ref{fig:contours1}.
}
\label{fig:west-contours}
\end{figure*}

As specified before, the West Coast Team has analyzed the mean over the 10 boxes in the high redshift bin at $z=0.61$, using the covariance on the mean.
Originally, for the purpose of parameter estimation, the Team presented the results up to $k_{\rm max}=0.12 \,\hMpci$
since this is the $k_{\rm max}$ at which the Team predicted the estimates to be still unbiased.
The marginalized posteriors for the cosmological parameters are shown in Fig.~\ref{fig:west-contours}, and best fit and means are listed in Table~\ref{tab:west-results}.
When the true results were revealed, it is found that $A_\mathrm{s}$ and $H_0$ lie within the $1$-$\sigma$ region of the estimates of the West Coast Team, and $\Omega_\mathrm{m}$ within the
$1.5$-$\sigma$ region.
$b_1$ is also correctly reproduced within the $1$-$\sigma$ interval. Additionally, one can see that the pre-unblinding results at $k_{\rm max}=0.14 \,\hMpci$, which however was not the $k_{\rm max}$ at which the Team anticipated to be most accurate, are even closer to the true values.

\begin{figure}[h!]
\begin{center}
\includegraphics[width=0.49\textwidth]{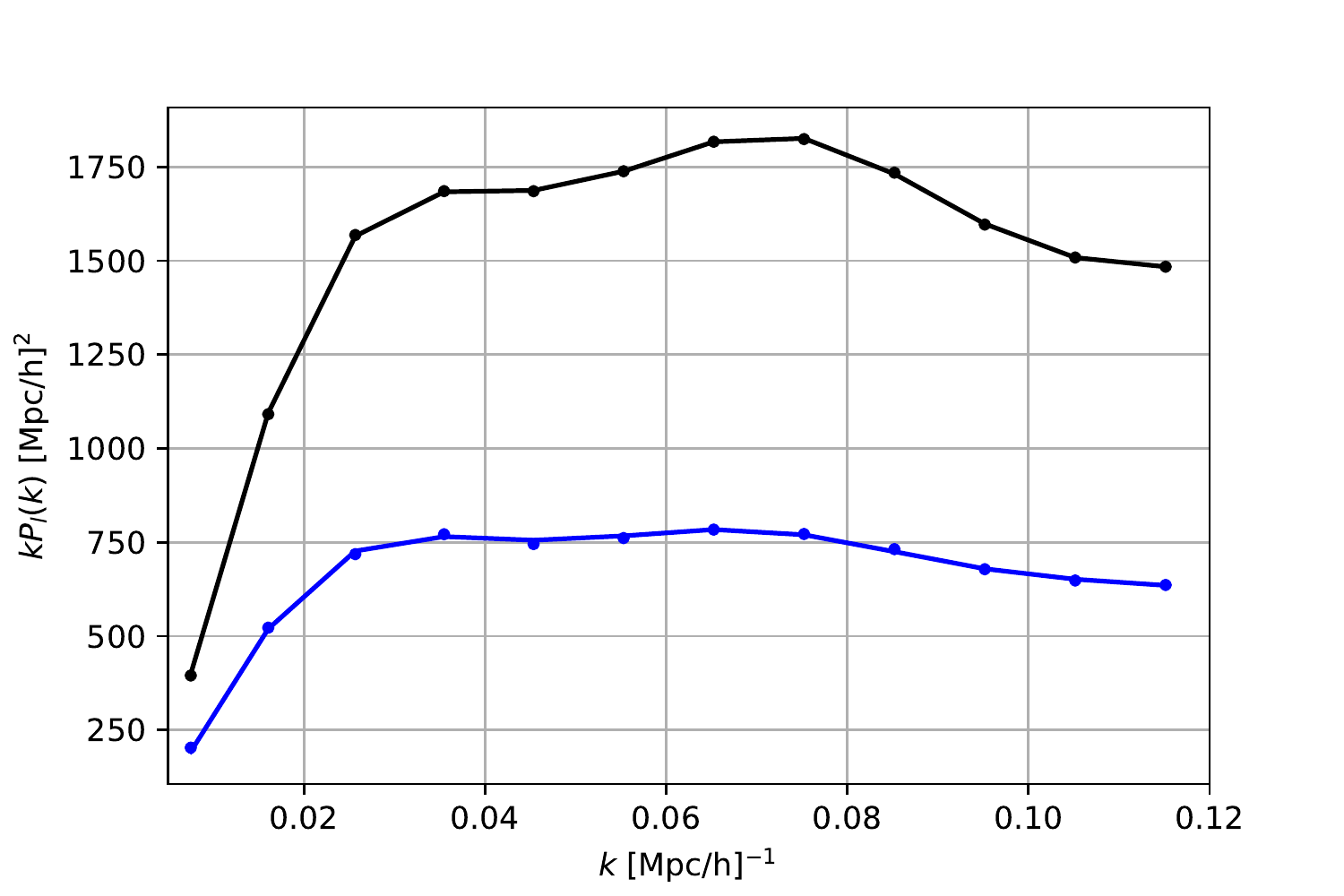}\\
\includegraphics[width=0.49\textwidth]{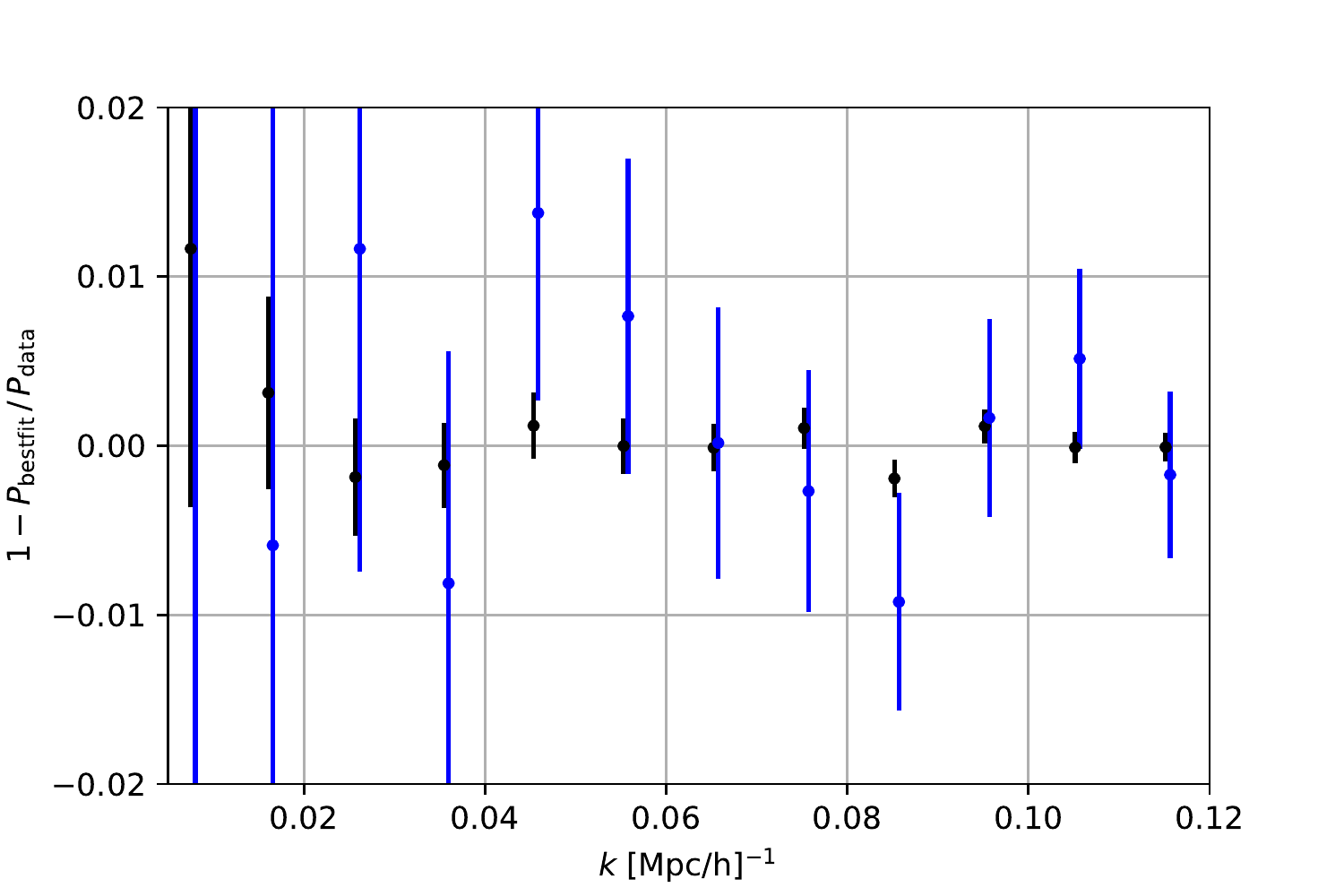}\\
\includegraphics[width=0.49\textwidth]{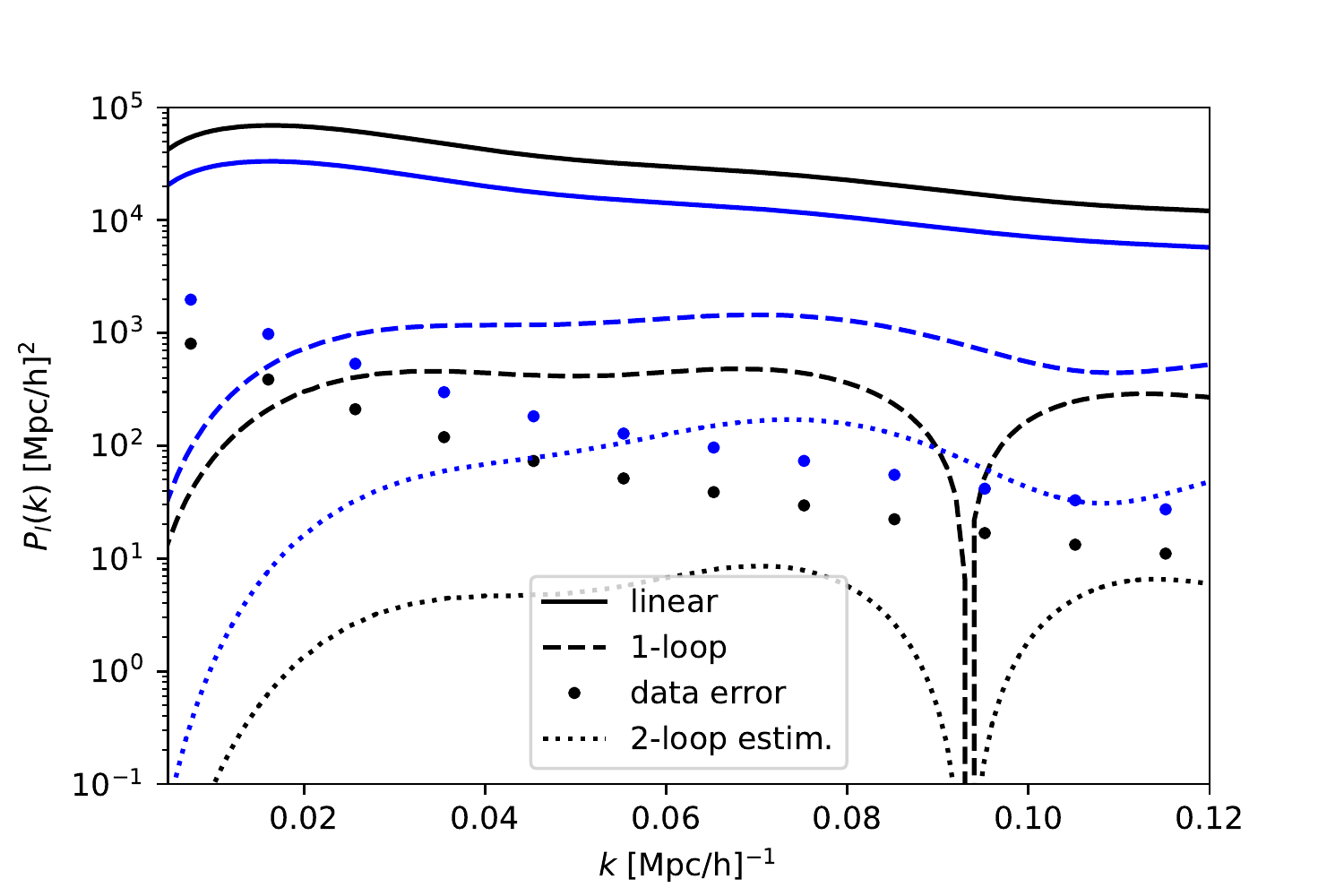}
\end{center}
\caption{\textit{Upper panel}: Comparison of the data for the monopole (black) and the quadrupole (blue) with the best-fit model obtained by the West Coast Team.
\textit{Middle panel}: Residuals for the monopole and the quadrupole for the best-fit model with the partially-marginalized Likelihood giving $-2 \log \mathcal{L}/{\rm dof}= 16/(24-6)$ for $k_{\rm max} = 0.12~\hMpci$.
\textit{Lower panel}:
Different contributions to the monopole and quadrupole power spectra. We plot just the absolute values, some terms are negative.}
\label{fig:west-bestfit}
\end{figure}

In Fig.~\ref{fig:west-bestfit} the Team shows that the data are well fitted by the theoretical model with the best-fit parameters, with
$-2 \log \mathcal{L}/\textrm{dof} = 16/(24-6)$,
corresponding to a very good $p$-value~\footnote{
Notice that the Likelihood of this team is not Gaussian.}.
In the lower panel, different contributions to the best fit power spectra are shown, to check the self-consistency of the perturbative expansion.
It is apparent that the one-loop term is safely less that $10\%$ of the linear one at all $k$'s.
In addition to the one-loop term, an estimate of the two-loop contribution, i.e. $P_{\rm 1-loop}^2/P_{\rm lin}$, is shown: clearly, at least for the quadrupole, this estimate is of the order of the error on the data at the highest $k$.
This is an additional indication that for roughly $k_{\rm max} \gtrsim 0.12\text{-}0.14 \, \hMpci$
the one-loop model will not be an accurate description of the data, and parameter estimation will suffer from theory systematics.

\emph{After unblinding}, the West Coast Team submitted additional results at $k_{\rm max}=0.14 , 0.16,0.18 , 0.20 \, \hMpci$.
This is because it was subsequently decided that it was interesting to explore the $k_{\rm max}$-dependence of the theory-systematic error.
In fact, though this has already been analyzed by the Team in both their original papers~\cite{DAmico:2019fhj,Colas:2019ret}, the challenge simulation is different and its volume larger.
At the higher $k_{\rm max}$'s, the Team performs the (analytical) marginalization over the additional $c_{\epsilon,\rm mono}$ parameter, with a Gaussian prior with $\sigma_{c_{\epsilon, \rm mono}}=2$.
The effect of adding this parameter is completely negligible at low $k_{\rm max}$: in fact, the Team chose to safely set it to zero for the original chains. Indeed one can check that the results are unchanged at low $k_{\rm max}$ when adding this parameter.
However, because of the small error bars of the simulation data, at higher $k_{\rm max}$ this parameter has to be added to the model.

The trend as a function of $k_{\rm max}$ is apparent from Fig.~\ref{fig:west-contours}.
$\Omega_\mathrm{m}$ and $H_0$ are well recovered up to $k_{\rm max} = 0.18 \, \hMpci$,
approximately within the $1\text{-}\sigma$ region, the estimate of clustering amplitude $A_\mathrm{s}$ starts to deviate significantly from the true value after $k_{\rm max} \gtrsim 0.14 \, \hMpci$.

\begin{table*}[ht]
\centering
\begin{tabular}{|l|c|c|}
 \hline
Param & best-fit & mean$\pm\sigma$  \\ \hline
$\Delta\Omega_\mathrm{m}/\Omega_\mathrm{m} \cdot 10^2$ &$1.3$ & $1.2_{-0.8}^{+0.8}$  \\
$\Delta h / h \cdot 10^2$ &$-0.7$ & $-0.6_{-0.6}^{+0.6}$ \\
$\Delta \ln (10^{10}A_{\rm s}) / \ln (10^{10}A_{\rm s}) \cdot 10^2$ &$0.1$ & $0.1_{-0.7}^{+0.7}$  \\
$\Delta b_1 /b_1\cdot 10^2$ &$0.8$ & $0.7_{-1.1}^{+1.0}$ \\
\hline
 \end{tabular}
  \caption{
  Similar to Table~\ref{tab:res}, but the
  results obtained by the West Coast Team for $k_{\rm max} = 0.12~\hMpci$
  at $z=0.61$. Only cosmological parameters and $b_1$ are shown.}
\label{tab:west-results}
\end{table*}

\subsection{Comparison of the two analyses}
\label{subsec:blind_summary}

\begin{figure}
\begin{center}
 \includegraphics[width=0.48\textwidth]
 {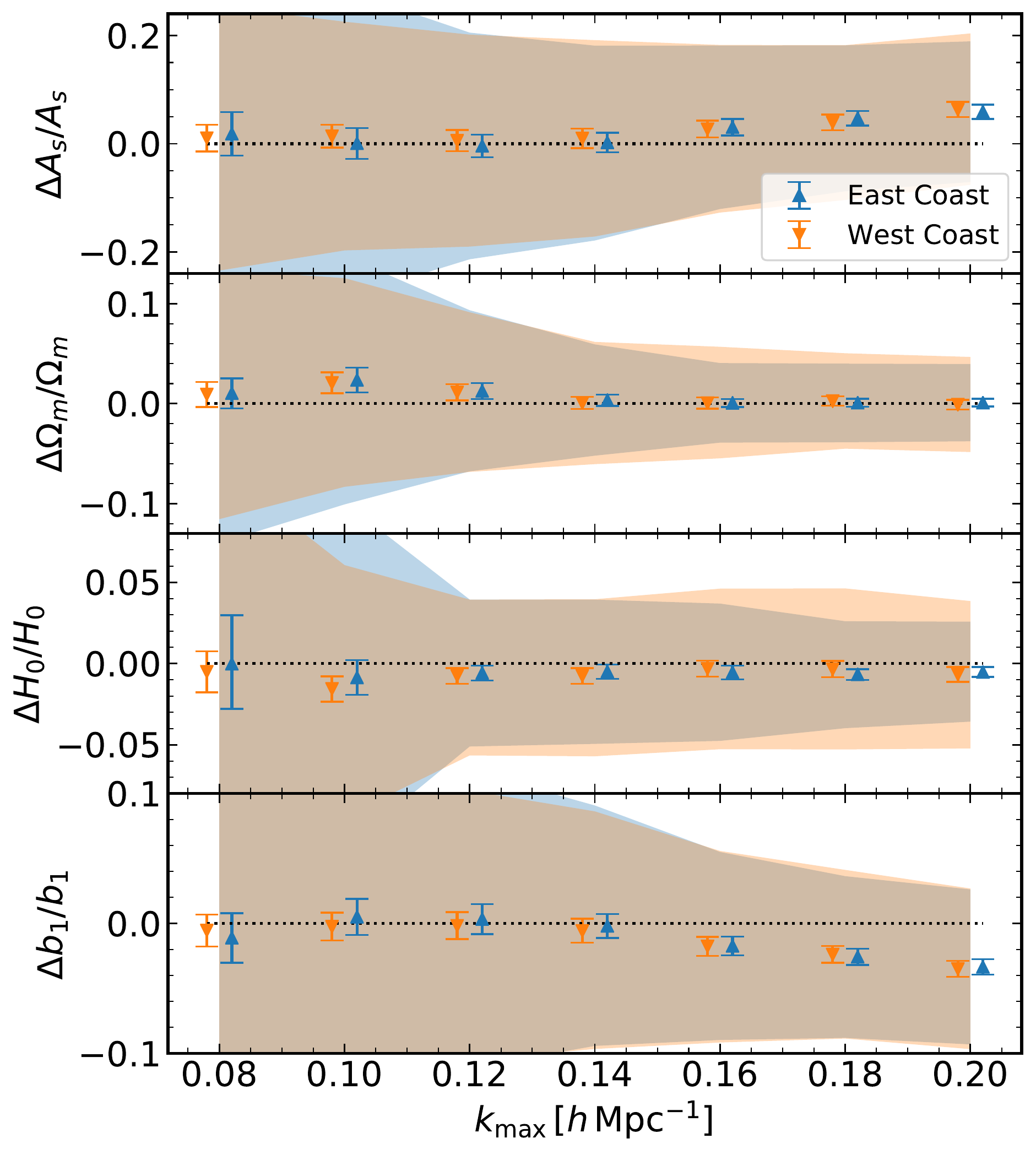}
 \end{center}
\caption{One dimensional
marginalized posterior distributions for the three main cosmological parameters as well as the linear bias parameter as a function of the maximum wavenumber $\kmax$ considered in the analysis. The 68\% credible intervals derived by the East and West Coast Team are shown respectively by the blue and red error bars with the mean marked by the upward and downward triangles, respectively.
Overplotted by the shaded regions are those scaled to the volume of SDSS DR12. The error bars are slightly shifted horizontally to avoid a heavy overlap.
}
\label{fig:1D_cosmo_main}
\end{figure}

\begin{figure*}
\begin{center}
 \includegraphics[width=0.48\textwidth]{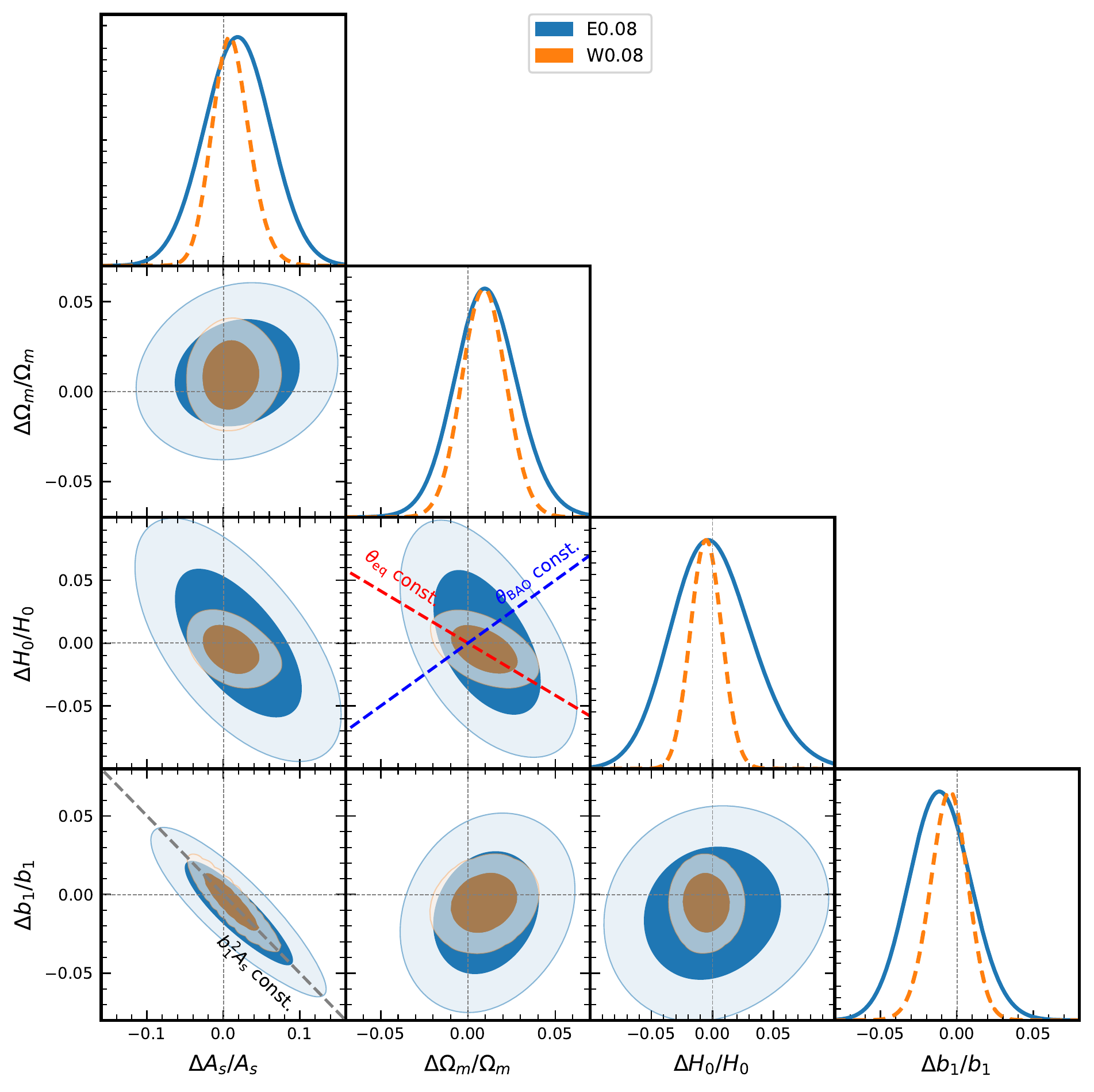}
 \includegraphics[width=0.48\textwidth]{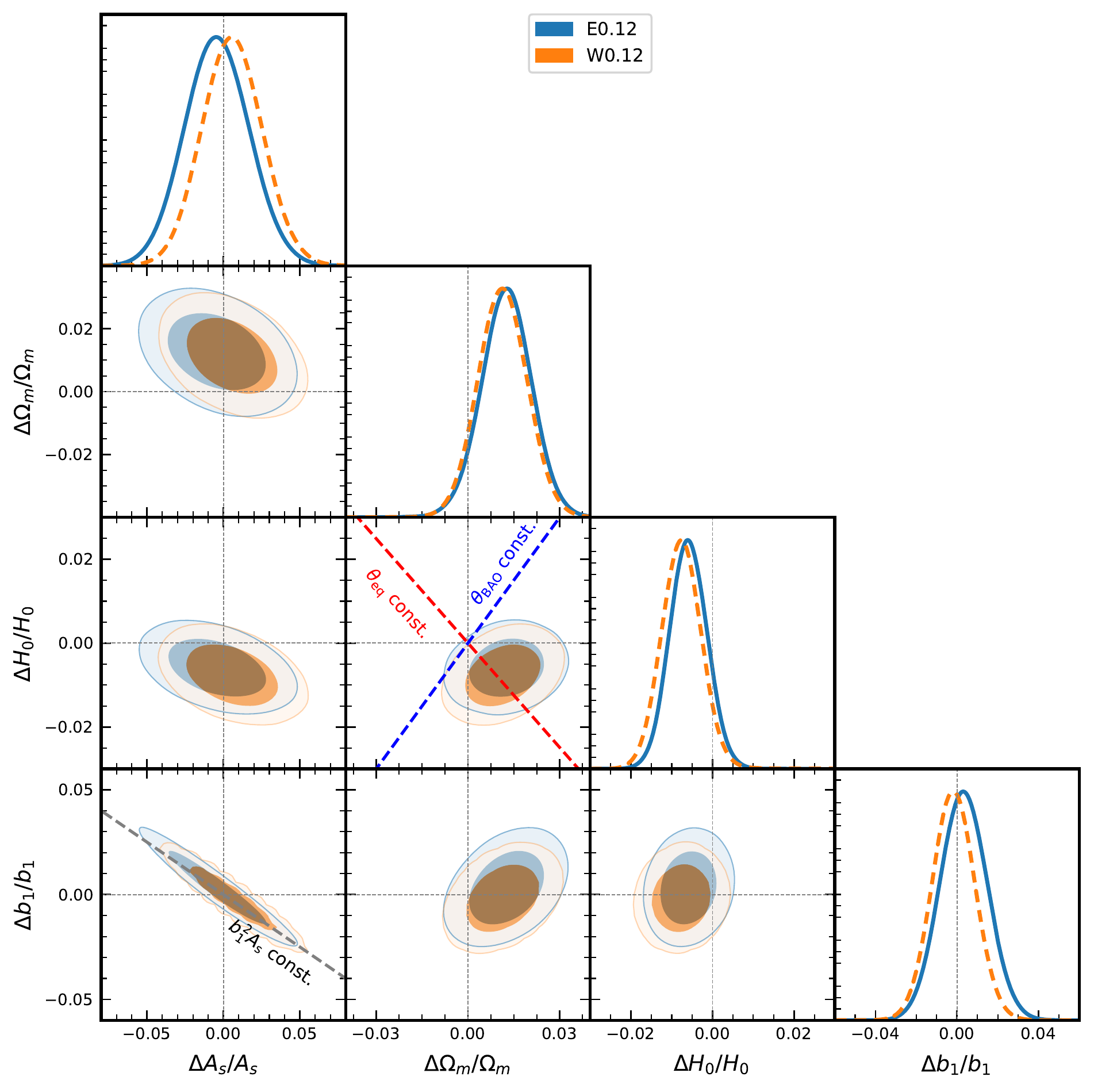}
 \includegraphics[width=0.48\textwidth]{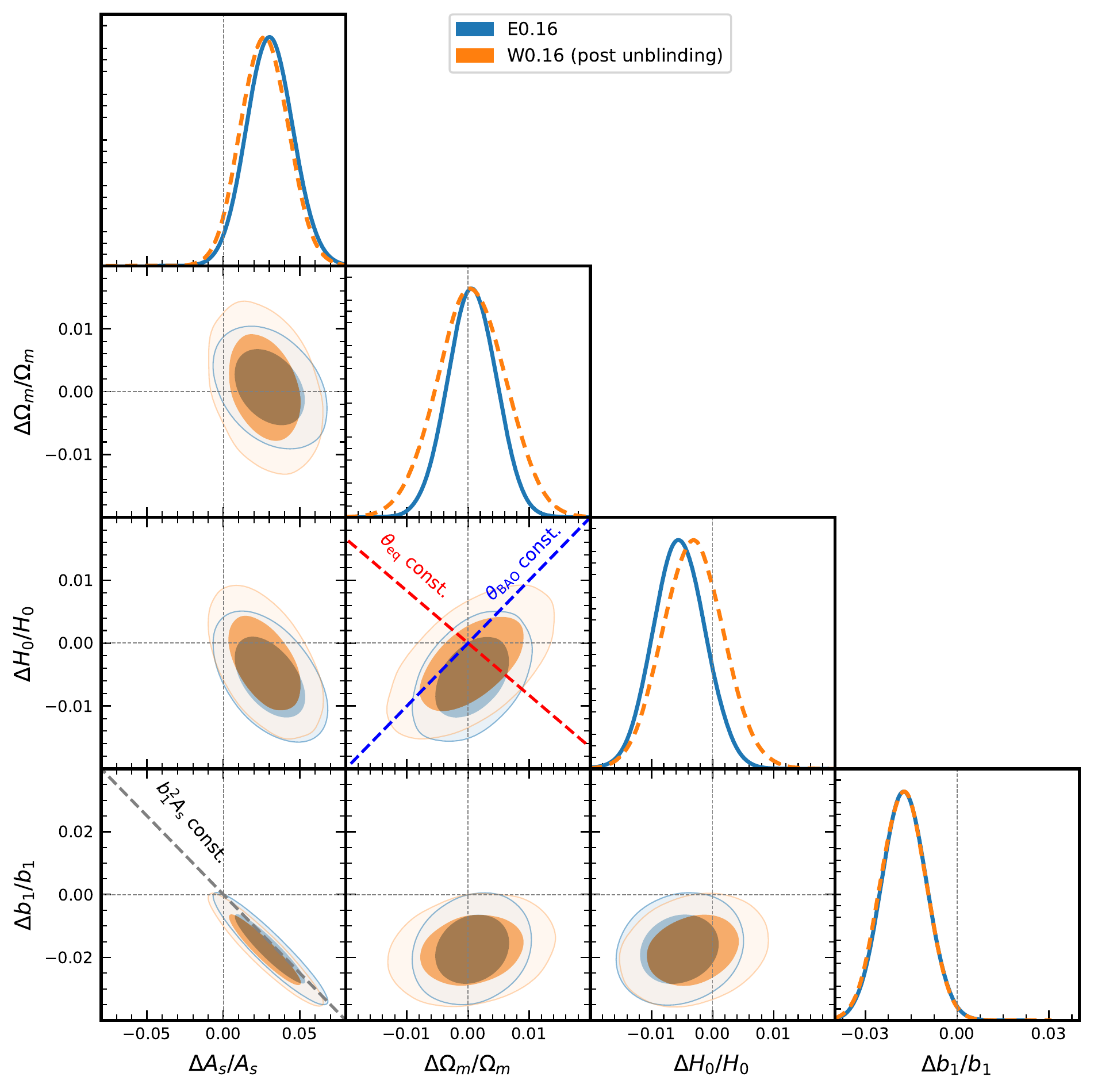}
 \includegraphics[width=0.48\textwidth]{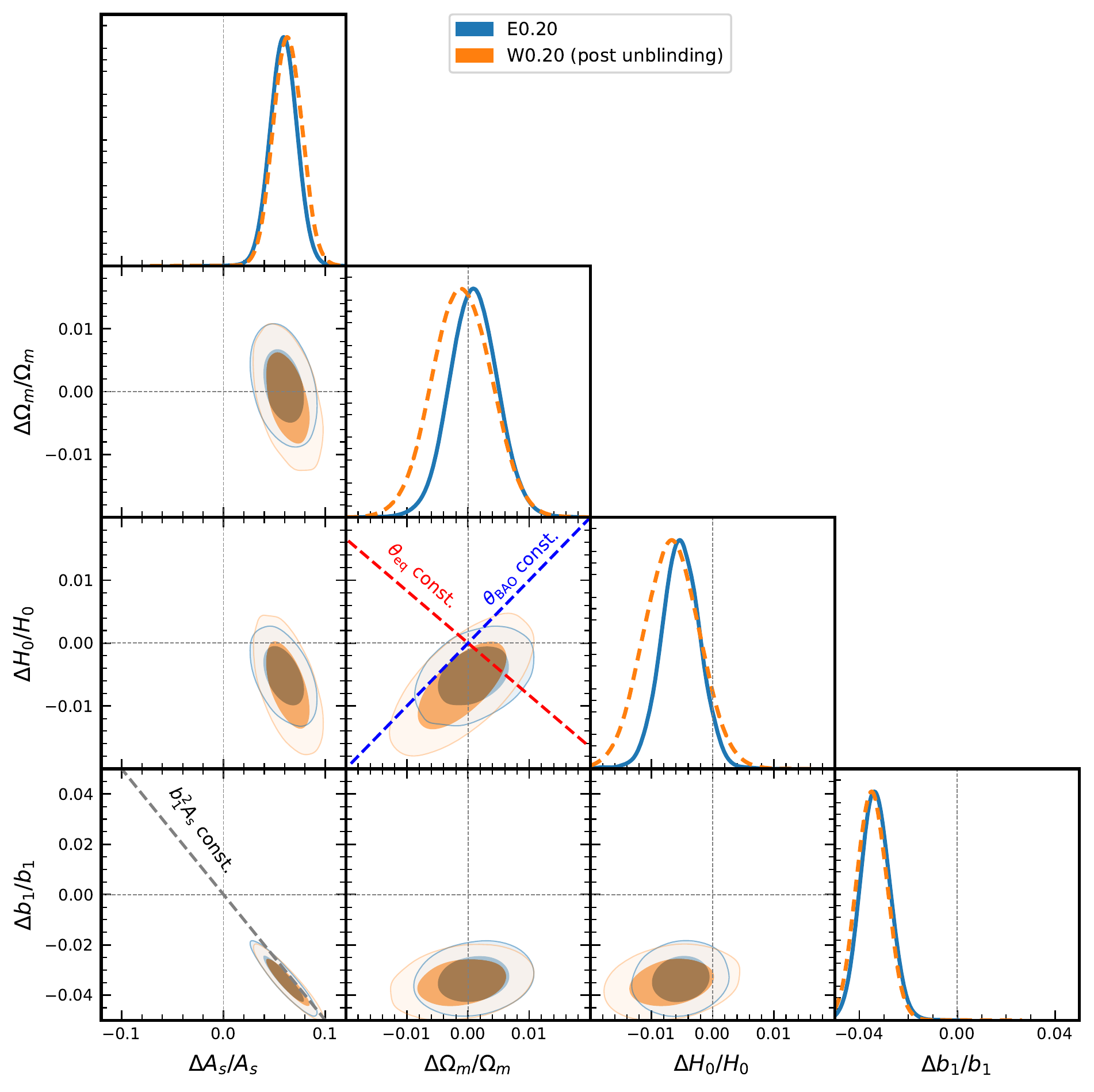}
\end{center}
\caption{Two dimensional marginal posterior distributions for the three main cosmological parameters and the linear bias parameter. The 68\% and 95\% credible intervals derived by the East and West Coast Team are shown respectively by the cyan and orange contours. The corresponding one dimensional marginal distributions are shown in the diagonal panels by the solid and dashed lines. The maximum wave number included in this analysis is $\kmax = 0.08$ (upper left), $0.12$ (upper right), $0.16$ (lower left) and $0.2\,\hMpci$ (lower right). Three degeneracy directions for some parameter combinations are also displayed in the contour panels by the thick dashed lines (see text for more detail).
}
\label{fig:2D_4panels}
\end{figure*}

So far we have presented the analyses done by two teams. We now compare the two and discuss how different model assumptions lead to the different cosmological-parameter constraints.

First, since the two teams employ different sets of cosmological parameters as the varied parameters, a direct comparison between
Figs.~\ref{fig:contours1} and \ref{fig:west-contours} is not very clear. We stick here instead to the parameter space $(\Omega_\mathrm{m}, H_0, A_\mathrm{s})$ to see the constraints. We first show in Fig.~\ref{fig:1D_cosmo_main} the one dimensional marginalized error on these parameters as a function of the maximum wavenumber, $k_\mathrm{max}$, used in the analysis. The $1$-$\sigma$ credible intervals by the East (West) Coast Team are shown by the upward (downward) triangles with error bars. Also shown by the shades are the same intervals but scaled for the SDSS BOSS DR12 according to the ratio of the simulated and the observed volume\footnote{We adopt the total volume of SDSS BOSS DR12, $5.7(\hiGpc)^3$, instead of that of CMASS2.}.

Overall, the ground truth values of the three cosmological parameters stay within or slightly off from the $1$-$\sigma$ interval up to $\kmax = 0.14~\hMpci$. The inferred primordial scalar amplitude, $A_\mathrm{s}$, in particular, is always within the interval up to this $\kmax$ from both teams.
On the one hand,
$A_\mathrm{s}$ starts to deviate from the ground truth in a systematic way with statistical significance
above this $\kmax$.
This is consistent with the expectation
that two-loop corrections become important at these scales.
On the other hand,~$H_0$ and $\Omega_\mathrm{m}$ stay roughly within $1\text{-}\sigma$ from
the true value all the way up to $k_{\rm max}=0.2~\hMpci$.
However, if one focuses
on the shaded regions corresponding to the statistical error from the actual BOSS survey, the ground truth values are always well within the $1$-$\sigma$ interval,
which justifies
the $k_{\rm max}$ choice
of the analyses from
the same teams in Refs.~\cite{DAmico:2019fhj,Ivanov:2019pdj,Colas:2019ret}.

While the size of the error bars shrinks towards higher $\kmax$, the gain is small after $\kmax \gtrsim 0.14~\hMpci$. This could be caused by the combination of two effects. Firstly, the relative contribution of the shot noise in the data covariance becomes important. Secondly, the EFT parameters controlling the nonlinear corrections
become
important in such a way that the additional information coming from small-scale modes mainly determines these parameters rather than the cosmological parameters. If one looks into the trend in the error bars more in detail, the results from the two teams are clearly different, especially when $\kmax \lesssim 0.1~\hMpci$, up to factor $\sim2$ smaller by the West Coast Team. This difference is driven
by the prior treatment.
The East Coast Team
had no priors on the chosen
set of nuisance parameters,
whereas the West Coast Team has
always kept the nuisance parameters
within physically-motivated
bounds.
Thus, the observed
difference of the results
between the two teams implies
that
on scales larger than
$0.1~\hMpci$ the data are not
good enough to break degeneracies
between the cosmological and
nuisance parameters.
These degeneracies get broken
at larger wavenumbers,
where the results of the two teams
agree regardless
of the nuisance parameters' priors.

Let us briefly
discuss some cosmological
implications of our blinded analysis.
The cosmological information
probed by redshift galaxy surveys
can be crudely divided
into four different categories:
\begin{itemize}
    \item Shape information.
    The shape of the galaxy power
    spectrum is controlled by
    the physical matter density
    $\omega_\mathrm{m}$. This parameter
    is measured from the data regardless of the choice of rulers such as $H_0$. $\omega_\mathrm{m}$ is extracted from the features
    of the power spectrum, such as
    the form of the BAO peaks,
    the baryonic suppression,
    the turnover, and the overall
     slope.
    \item Distance information, mainly encoded through the volume-average distance\footnote{It is defined as $D_\mathrm{V}(z)=\left(z(1+z)^2D^2_\mathrm{A}(z)/H(z)\right)^{1/3}$, where $D_\mathrm{A}(z)=\frac{1}{1+z}\int_0^z \frac{dz'}{H(z')}$ and $H^2(z)=H^2_0(\Omega_{\rm m}(1+z)^3+1-\Omega_\mathrm{m})$ in flat $\Lambda$CDM.} $D_\mathrm{V}(z)$.
    This parameter, essentially, controls the freedom
    to shift the power spectrum along
    the $k$ axis.
    In the flat $\Lambda$CDM framework this distance depends only on two cosmological parameters, $\omega_\mathrm{m}$ and $H_0$. Since $\omega_\mathrm{m}$
    is measured from the shape, the constraint on $D_\mathrm{V}$ translates directly into a constraint on $H_0$.
    Note that $\Omega_\mathrm{m}$ in this picture can be seen as a parameter derived from a  combination of the shape and distance parameters.
    \item Redshift space distortions. Observing galaxies in redshift
    space allows one to measure unbiased
    rms velocity fluctuation $f\sigma_8(z)=f(z)D_+(z)\sigma_8$. In $\Lambda$CDM $D_+$ and $f$ depend only on $\Omega_{\rm m}$, which
    is constrained from the shape
    and the distance.
    This way the RSD measurements constrain
    directly $A_\mathrm{s}$.
    \item The Alcock-Paczynski geometric distance information. The AP effect allows one to measure the combination $H(z)D_\mathrm{A}(z)$.
    However, in $\Lambda$CDM this combination is a slow function
    of cosmological parameters
    at small redshifts. Thus, it does not contribute significantly to the overall constraints on $\Omega_\mathrm{m}$, see Ref.~\cite{Ivanov:2019pdj} for more detail.
\end{itemize}

We can see in Fig.~\ref{fig:2D_4panels} that indeed our results are fully in line with these theoretical expectations.
First, let us focus on the two-dimensional posterior in the $(\Omega_\mathrm{m}$--$H_0)$ plane. The change in the degeneracy direction is observed to rotate with increasing the maximum wavenumber. When $\kmax=0.08\,\hMpci$, $\Omega_\mathrm{m}$ and $H_0$ are negatively correlated. At the other end, the correlation turns to be a positive one for $\kmax=0.16$ and $0.2\,\hMpci$. We can interpret this as the outcome of the change in the relative importance of the BAO feature. Although the first BAO peak is already included at $\kmax=0.08\,\hMpci$, the dominant constraint is coming from the overall shape information, e.g., the matter-radiation equality scale ($\theta_\mathrm{eq} = 1/(k_\mathrm{eq}D_\mathrm{V}) \propto \Omega_\mathrm{m}^{-0.83}h^{-1}$, where $k_\mathrm{eq}$ denotes the equality wavenumber) at this maximum wavenumber. Indeed, the contours from the two teams are roughly oriented along this direction depicted by the red dashed line. At $\kmax=0.2\,\hMpci$, as we can clearly see the BAO feature up to the third peak (see Fig.~\ref{fig:multipole}), the BAO scale (the blue dashed line in Fig.~\ref{fig:2D_4panels}: $\theta_\mathrm{BAO} = r_\mathrm{s}/D_\mathrm{V}$ with the sound horizon scale $r_\mathrm{s}$) plays a more significant role. The measurement of the relative location of these two characteristic scales allow us to determine the physical density $\omega_\mathrm{m} = \Omega_\mathrm{m} h^2$, and together with the distance measurement through cosmology dependence of the redshift-distance conversion (i.e., a measurement of $D_\mathrm{V}$), we can break the degeneracy between $\Omega_\mathrm{m}$ and $H_0$.

\begin{figure}[h!]
\begin{center}
 \includegraphics[width=0.48\textwidth]{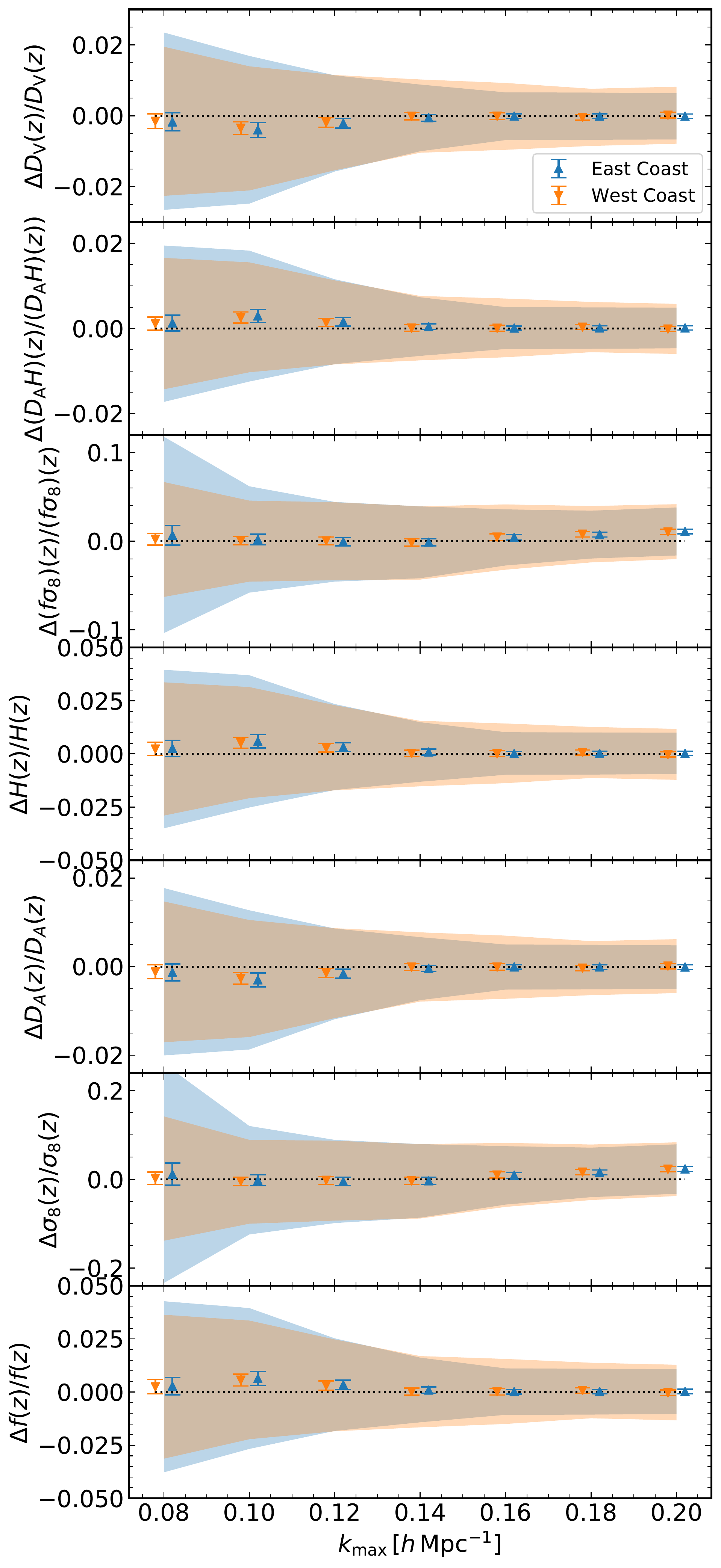}
 \end{center}
\caption{One dimensional marginalized
posterior distributions of {\it derived} parameters for flat $\Lambda$CDM model
as a function of the maximum wavenumber included in the analysis, $k_\mathrm{max}$. The fractional error is shown with the uncertainty in $H_0$
that
is kept in the unit for the distance parameters (i.e., $D_\mathrm{A}$ is expressed in $\hiMpc$ and $H$ is in $\hMpci$).
}
\label{fig:1D_other}
\end{figure}

Once $D_\mathrm{V}(z)$ and $\omega_{\rm m}$ are fixed, the other parameters such as the distance parameters, $H(z)$, $D_\mathrm{A}(z)$ (with $h$ kept in the unit as $\hMpci$ or $\hiMpc$) or the growth parameter, $f(z)$, are merely dependent parameters fully determined by $\Omega_\mathrm{m}$ given that we stick to the flat $\Lambda$CDM cosmology. Had we fitted the data with a more general expansion model, e.g.~dynamical dark energy or modified gravity models, the posterior distribution of these parameters would have been different.
These parameters extracted from our MCMC chains, together with some other useful parameters, are displayed in Fig.~\ref{fig:1D_other}.

\begin{figure}
\begin{center}
 \includegraphics[width=0.48\textwidth]{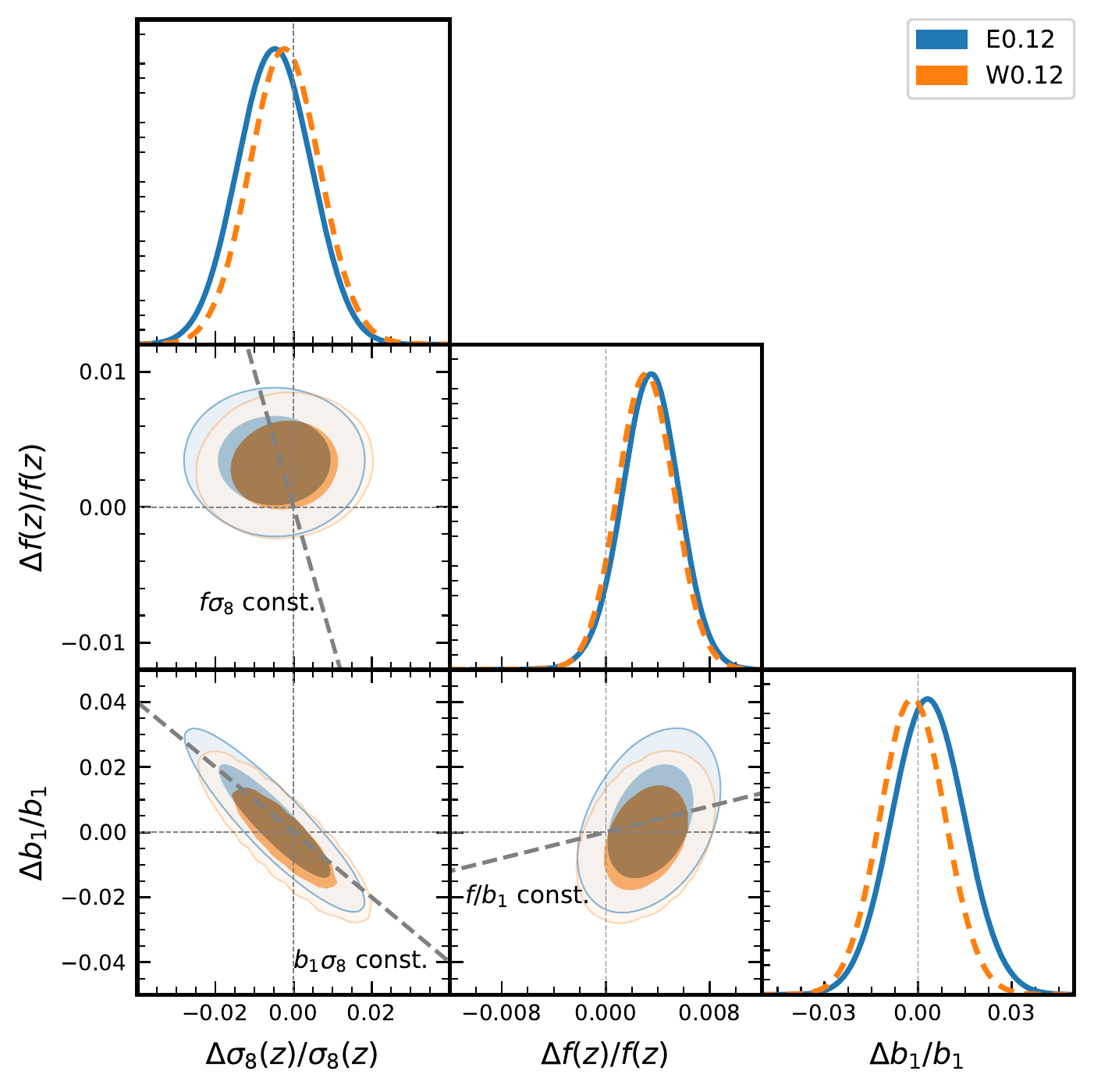}
 \end{center}
\caption{Two dimensional marginalized posterior
distributions
for amplitude-related parameters
relevant for the RSD measurement from the analyses at $\kmax=0.12\,\hMpci$. The expected degeneracy directions, $f\sigma_8$, $b_1\sigma_8$ or $f/b_1$ is constant, expected from linear RSD measurements are shown by the dashed lines.
Note that $f(z)$ and $\sigma_8(z)$ are derived parameters
fully fixed once $\Omega_\mathrm{m}$ and $A_\mathrm{s}$ are given within the flat $\Lambda$CDM model.
}
\label{fig:2D_amplitude}
\end{figure}

Apart from the shape-related parameters, the determination of the amplitude parameter is of interest. We can see in Fig.~\ref{fig:2D_4panels} that the posterior of the amplitude parameter, $A_\mathrm{s}$, is strongly correlated with the linear bias parameter $b_1$. To understand this more clearly, we show the constraints on the parameters relevant for the measurement of RSD (the one-dimensional and the two-dimensional marginalized posterior in Figs.~\ref{fig:1D_other} and \ref{fig:2D_amplitude}, respectively). In the two-dimensional contour plot, we can see that the amplitude parameter scaled to the redshift of the survey volume, $\sigma_8(z) = [D_+(z)/D_+(z=0)]\,\sigma_8$, is strongly degenerate with the linear bias parameter, $b_1$, just as we have seen for $A_\mathrm{s}$ and $b_1$.
In fact, they are expected to be fully degenerate
in the absence of RSD information in linear theory. We can also see in Fig.~\ref{fig:1D_cosmo_main}
that $b_1$ starts to depend weakly on $\kmax$ above
$\sim 0.14\,\hMpci$ with statistical significance, and a similar departure from the ground truth value happens at the same place but to the opposite direction in $\sigma_8(z)$ as shown in Fig.~\ref{fig:1D_other}.
The other famous degeneracy directions, $f\sigma_8$ or $f/b$, which are the direct observables from linear RSD, do not appear in our contours in Fig.~\ref{fig:2D_amplitude}. This is again due to the fact that the flat $\Lambda$CDM assumption makes $f$ a dependent variable fully determined by $\Omega_\mathrm{m}$. What we see here is that the constraint on $\Omega_\mathrm{m}$ through the shape and distance measurement discussed above, combined with the measurement of $f\sigma_8$ from RSD, allows us to constrain $\sigma_8$ (and thus $A_\mathrm{s}$) directly.

\section{Conclusion and outlook}
\label{sec:conclusion}

In this paper we have presented results of the blinded
cosmology
challenge
initiated
to test theoretical models for redshift-space
galaxy clustering.
The task was to assess whether the theoretical model, here EFTofLSS, can recover the blinded cosmological parameters in N-body simulation from the mock data of redshift-space power spectrum multipoles for BOSS-like galaxies.
The sufficiently large volume,
dynamical range and high resolution
of the challenge simulation
allow
one to pin down any
potential
inaccuracy of
theoretical modeling, compared to the statistical errors for the BOSS-like survey.

The simulations
were run by a team (``Japan Team'') that
kept the true parameters
in secret. The
mock data
were analyzed by two
other independent teams
(``East Coast Team"
and ``West Coast Team'')
who volunteered to participate in the challenge.
The rule of the challenge
is that the true parameters can be unblinded
only when the analyzing teams submit their final results to the simulation team.
All the three teams agreed that the submitted results be presented in this paper, without any change, after the unblinding.

Both analyzing teams
used the same theoretical model
based on
the effective field theory of large-scale structure.
However,
there exist some nontrivial differences, whose impact on the final cosmological inference should be tested quantitatively with care.
The corresponding pipelines were
the ones applied to the real BOSS data in Refs.~\cite{DAmico:2019fhj,Ivanov:2019pdj,Colas:2019ret}.
We have discussed in detail methodological and technical differences between these two pipelines.
Despite these
differences,
both teams have successfully recovered the true cosmological parameters within
expected statistical error bars.
This suggests that
perturbation theory,
once consistently implemented,
can be used
as a standard tool
for unbiased estimation
of cosmological parameters from galaxy surveys.

The enormously large total simulation volume used in the challenge helped to assess systematic error due to the incomplete theoretical modeling by suppressing statistical error to a level much lower than the current surveys.
The biased cosmological inference beyond the reported maximum wavenumber used for the challenge, $\kmax=0.12\,\hMpci$, consistently determined by both teams, indicates the typical systematic error one can make from actual surveys with much smaller observed volume (see, e.g., Fig.~\ref{fig:1D_cosmo_main} up to $\kmax=0.2\,\hMpci$). For instance, the analyses of SDSS BOSS galaxies by Refs.~\cite{DAmico:2019fhj,Ivanov:2019pdj,Colas:2019ret} adopt $\kmax$ around $0.2\,\hMpci$ (0.18 to 0.25 depending on the paper and the redshift bin of the galaxy sample). While the detailed choice of varied cosmological parameters as well as the way to combine with CMB constraints are different from what is presented here, one can make a reasonable guess on the potential systematic biases on the inferred cosmological parameters of these papers out of our results.

Out of the three cosmological parameters that we considered here, the scalar amplitude parameter, $A_\mathrm{s}$, is most severely biased beyond $\kmax=0.12\hMpci$, reaching $\sim 4\%$\footnote{We estimate this theory sytematic error as the distance from the truth of the $1 \sigma$ region of the posterior, as done in~\cite{DAmico:2019fhj}.} at $\kmax=0.2\,\hMpci$, while the two other parameters, $\Omega_\mathrm{m}$ and $H_0$, are fairly unbiased even when the EFT template starts to fail. This indicates that the latter two are mostly constrained through the shape of the spectrum (mostly the distinctive BAO feature)
The situation should be the same in actual observational data analyses such as the one listed above. Although the precise value of the detected parameter bias on $A_\mathrm{s}$ can depend on the detail in the halo-galaxy connection mainly through the uncertainty in the strength of the redshift-space distortions,
it is assuring to observe that our worst case value of $4\%$ is still below the statistical error from Refs.~\cite{DAmico:2019fhj,Ivanov:2019pdj,Colas:2019ret}, which are $12\%$ to $19\%$ ($68\%$ C.L.) depending on the paper.
Future experiments with even larger survey volume and higher galaxy number density will allow us to lower these uncertainties and in that case one have to be more careful on the parameter bias due to the model inaccuracy, either by lowering $\kmax$ or by improving the model itself. We investigate the parameter constraints for a hypothetical survey with the volume of Dark Energy Spectroscopic Instrument (DESI: \cite{DESI}) in Appendix~\ref{app:desi}.

We are currently exploring
a number of various post-blinded research
directions.
The first one includes a thorough investigation of the information content of redshift galaxy surveys.
Second, it would be curious to see how much
the
$k_{\rm max}$ value
where one-loop perturbation theory
breaks down depends
on the properties
of the galaxy population,
i.e. assembly bias or
satellite fraction.
Third, it will be interesting to see how well perturbation theory
performs for other observables, e.g.
the
galaxy-galaxy weak lensing
or the redshift-space bispectrum.
These research avenues are
left for future work.

We have presented the results
obtained by analyzing teams in the way such that the
true parameters are still
blinded to the readers.
This is done in case
some
other researchers would
like to test their theory models on the challenge spectra.
All challenge data are available online at
\url{http://www2.yukawa.kyoto-u.ac.jp/~takahiro.nishimichi/data/PTchallenge/}.
We encourage all groups
working on galaxy clustering analysis to
participate in the
challenge.


\acknowledgements{TN, LS, MT and MZ acknowledge a warm hospitality of the BCCP-IAS workshop ``The Nonlinear Universe 2018'' held at Smartno, Slovenia, where this work was initiated.
This work is supported in part by World Premier International Research Center Initiative (WPI Initiative), MEXT, Japan, and
by MEXT/JSPS KAKENHI Grant Numbers JP17K14273 (TN),
JP15H05887 (MT), JP15H05893 (MT),  JP15K21733 (MT), and JP19H00677 (TN, MT).
TN also acknowledges financial support from Japan Science and Technology Agency (JST) CREST Grant Number JPMJCR1414 and by JST AIP Acceleration Research Grant Number JP20317829, Japan. Numerical computations were carried out on Cray XC50 at Center for Computational Astrophysics, National Astronomical Observatory of Japan. GDA is partially supported by Simons Foundation Origins of the Universe program (Modern Inflationary Cosmology collaboration).
LS is partially supported by Simons Foundation Origins of the Universe program (Modern Inflationary Cosmology collaboration) and by NSF award 1720397.
MZ is supported by NSF grants AST1409709, PHY-1820775 the Canadian
Institute for Advanced Research (CIFAR) program on
Gravity and the Extreme Universe and the Simons Foundation Modern Inflationary Cosmology initiative.
MI is partially supported by the Simons Foundation’s Origins of the Universe program and by the RFBR grant 20-02-00982~A.
}

\appendix
 \section{\label{sec:galaxykernels} Galaxy kernels}
 The explicit expressions for the galaxy kernels appearing in the one-loop power spectrum are given here (see for a derivation \cite{Perko:2016puo}):
\begin{align}\label{eq:redshift_kernels}\nonumber
    Z_1(\q_1) & = K_1(\q_1) +f\mu_1^2 G_1(\q_1) = b_1 + f\mu_1^2,\\ \nonumber
    Z_2(\q_1,\q_2,\mu) & = K_2(\q_1,\q_2) +f\mu_{12}^2 G_2(\q_1,\q_2)\\ \nonumber
    &+ \, \frac{1}{2}f \mu q \left( \frac{\mu_2}{q_2}G_1(\q_2) Z_1(\q_1) + \text{perm.} \right),\\ \nonumber
    Z_3(\q_1,\q_2,\q_3,\mu) & = K_3(\q_1,\q_2,\q_3) + f\mu_{123}^2 G_3(\q_1,\q_2,\q_3) \nonumber\\ \nonumber
    & + \frac{1}{3}f\mu q \left(\frac{\mu_3}{q_3} G_1(\q_3) Z_2(\q_1,\q_2,\mu_{123})  \right.\\  &\left.+\frac{\mu_{23}}{q_{23}}G_2(\q_2,\q_3)Z_1(\q_1)+ \text{cyc.}\right),
\end{align}
where here $\mu= \q \cdot \hat{\z}/q$, $\q = \q_1 + \dots +\q_n$, and $\mu_{i_1\ldots  i_n} = \q_{i_1\ldots  i_n} \cdot \hat{\z}/q_{i_1\ldots  i_n}$, $\q_{i_1 \dots i_m}=\q_{i_1} + \dots +\q_{i_m}$, with $\hat{\z}$ being the unit vector in the direction of the line of sight, and $n$ is the order of the kernel $Z_n$. $K_i$ and $G_i$ are the galaxy density and velocity kernels, respectively.
We choose to work in the basis of descendants (this is the first complete set of bias coefficient for LSS, established in \cite{Senatore:2014eva,Angulo:2015eqa} and with some typos corrected in \cite{Fujita:2016dne}; see~\cite{Senatore:2014eva,Angulo:2015eqa} for connection to former bases of bias coefficients, as for example~\cite{McDonald:2009dh}). Notice that while the new terms introduced in \cite{Senatore:2014eva} happen to be degenerate with the standard bias terms at one-loop order, this will not be {the} case anymore once one goes to higher orders. For the one-loop power spectrum, all kernels can be described with 4 bias parameters~$b_i$.

The first and second order galaxy density kernel are:
 \begin{align}
     K_1 & = b_1, \\ \nonumber
     K_2(\q_1,\q_2) & = b_1\frac{\q_1\cdot \q_2}{q_1^2}+ b_2\left( F_2(\q_1,\q_2)- \frac{\q_1\cdot \q_2}{q_1^2} \right) \\ \nonumber
     &+ b_4 + \text{perm.} \ .
 \end{align}
The galaxy velocity kernels $G_n$ are simply the standard perturbation theory ones since the galaxy velocity field follows the dark matter velocity field, up to higher-derivative terms  which are degenerate with other counterterms that appear in the renormalization of the redshift space expression  (see e.g. \cite{Bernardeau:2001qr} for the expressions of $F_n$ and $G_n$).

The third-order galaxy density kernel has a much more involved expression. However, for the one-loop calculation, degeneracies appear in the one-loop diagram obtained from $\langle\delta^{(3)}\delta^{(1)}\rangle$, when UV-divergences are removed and the integral over the angular coordinates is performed, leading to the following simple expression:
\begin{align}
    K_3(k,q) & = \frac{b_1}{504 k^3 q^3}\left( -38 k^5q + 48 k^3 q^3 - 18 kq^5 \right. \\ \nonumber
    &\left.\qquad\qquad\quad+ 9 (k^2-q^2)^3\log \left[\frac{k-q}{k+q}\right] \right) \nonumber \\\nonumber
    &+ \frac{b_3}{756 k^3 q^5} \left( 2kq(k^2+q^2)(3k^4-14k^2q^2+3q^4)\right.\\ \nonumber
    &\left.\qquad\qquad \quad+3(k^2-q^2)^4 \log \left[\frac{k-q}{k+q}\right]  \right).
\end{align}

\section{Post-unblinding analyses}
\label{app:post}
While both of the analysis teams have worked on a specific bias parameterization in the main text, it is worth exploring different options for better understanding and to make a better connection to some of the recent works on observational data.
This Appendix presents two such possibilities.

\begin{figure*}[ht]
\begin{center}
\includegraphics[width=0.49\textwidth]{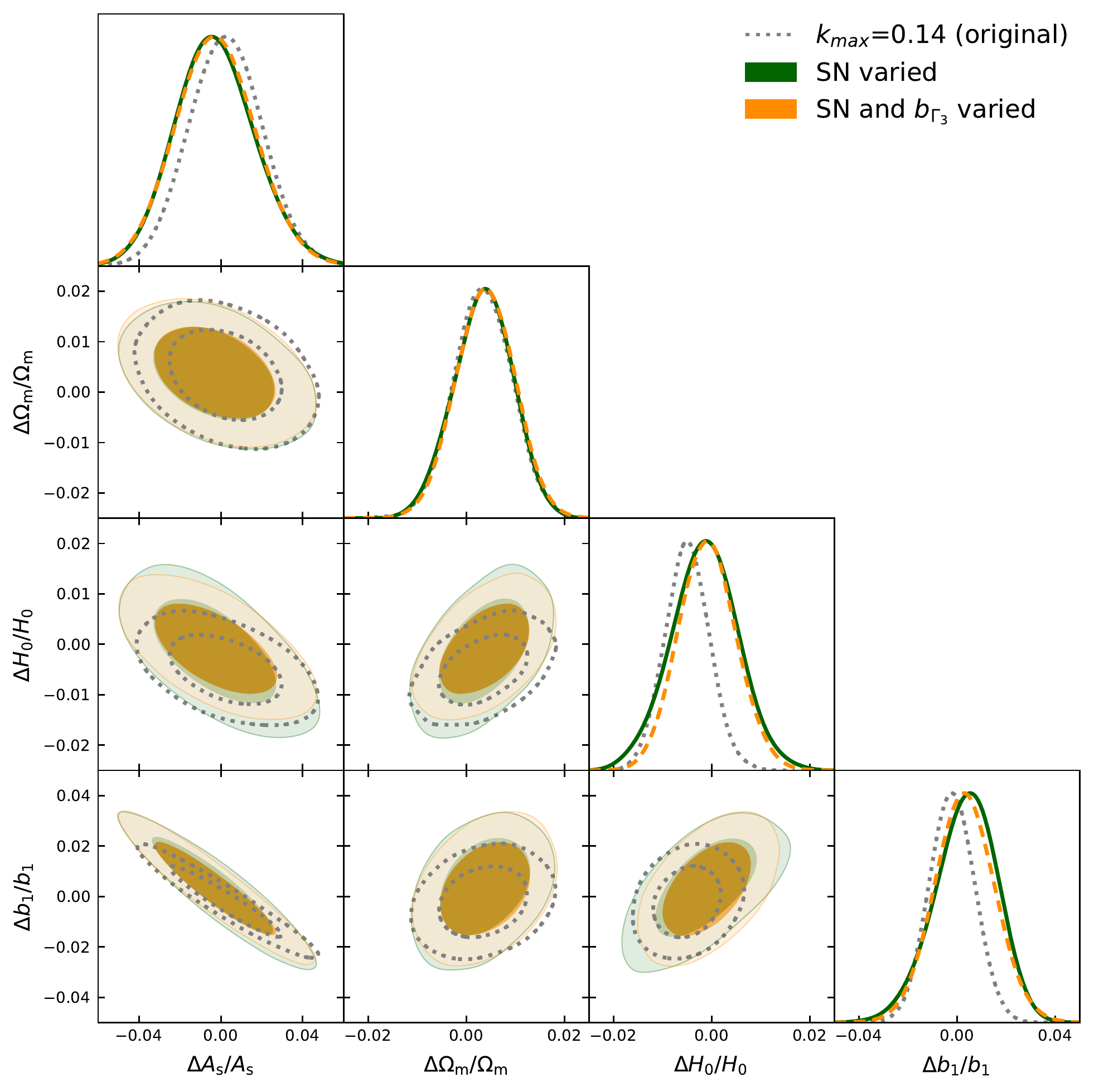}
\includegraphics[width=0.49\textwidth]{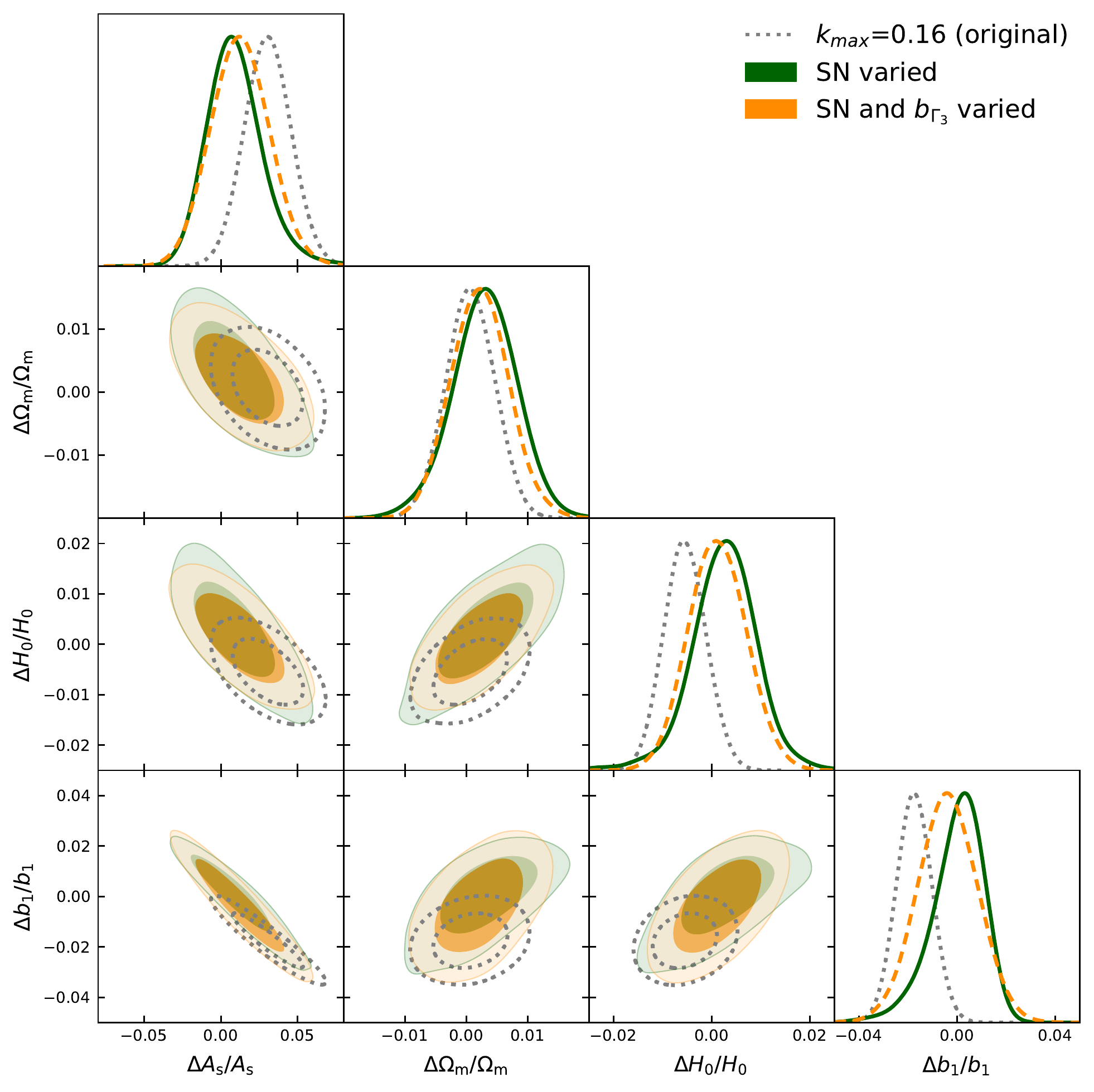}
\includegraphics[width=0.49\textwidth]{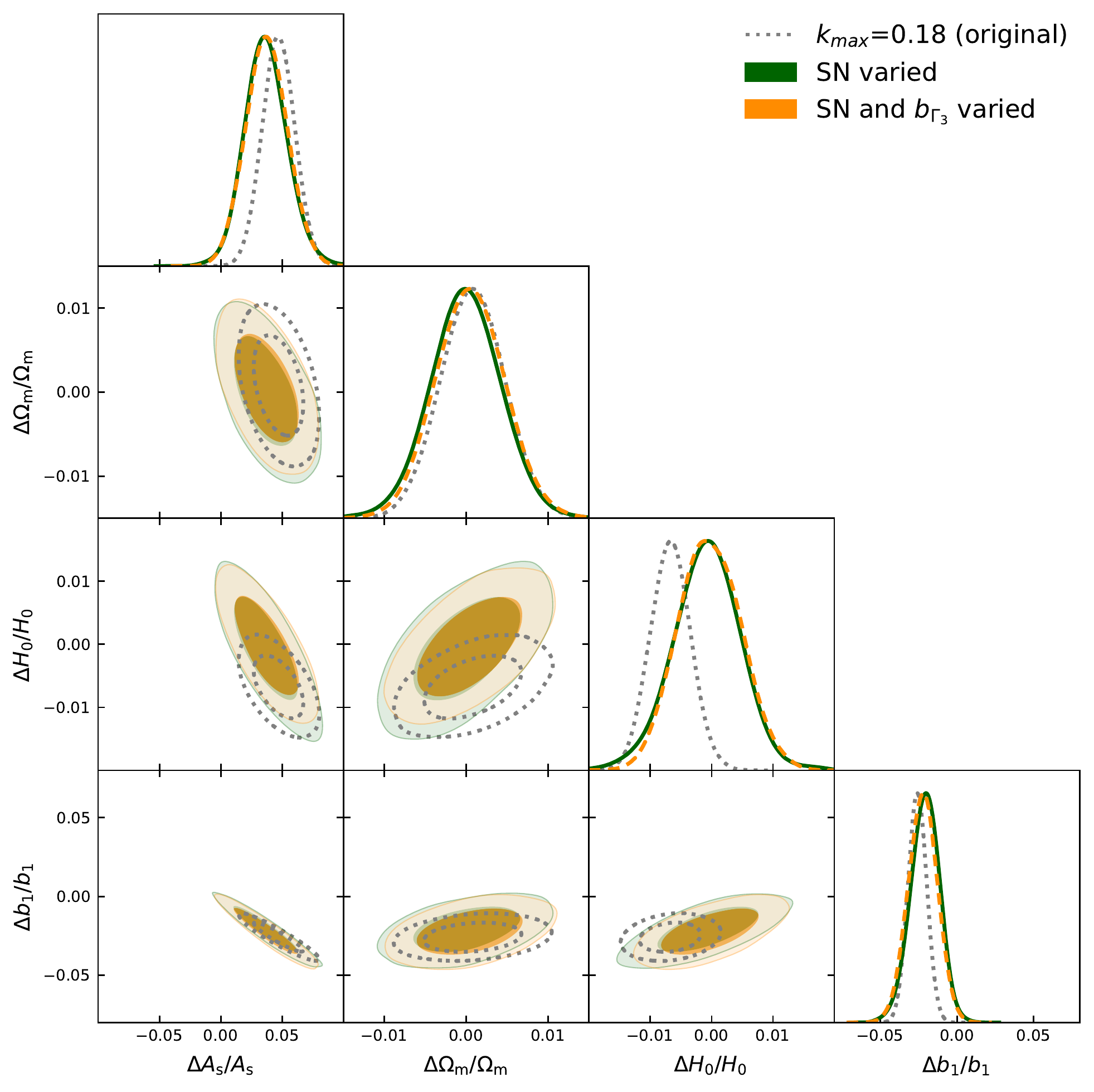}
\includegraphics[width=0.49\textwidth]{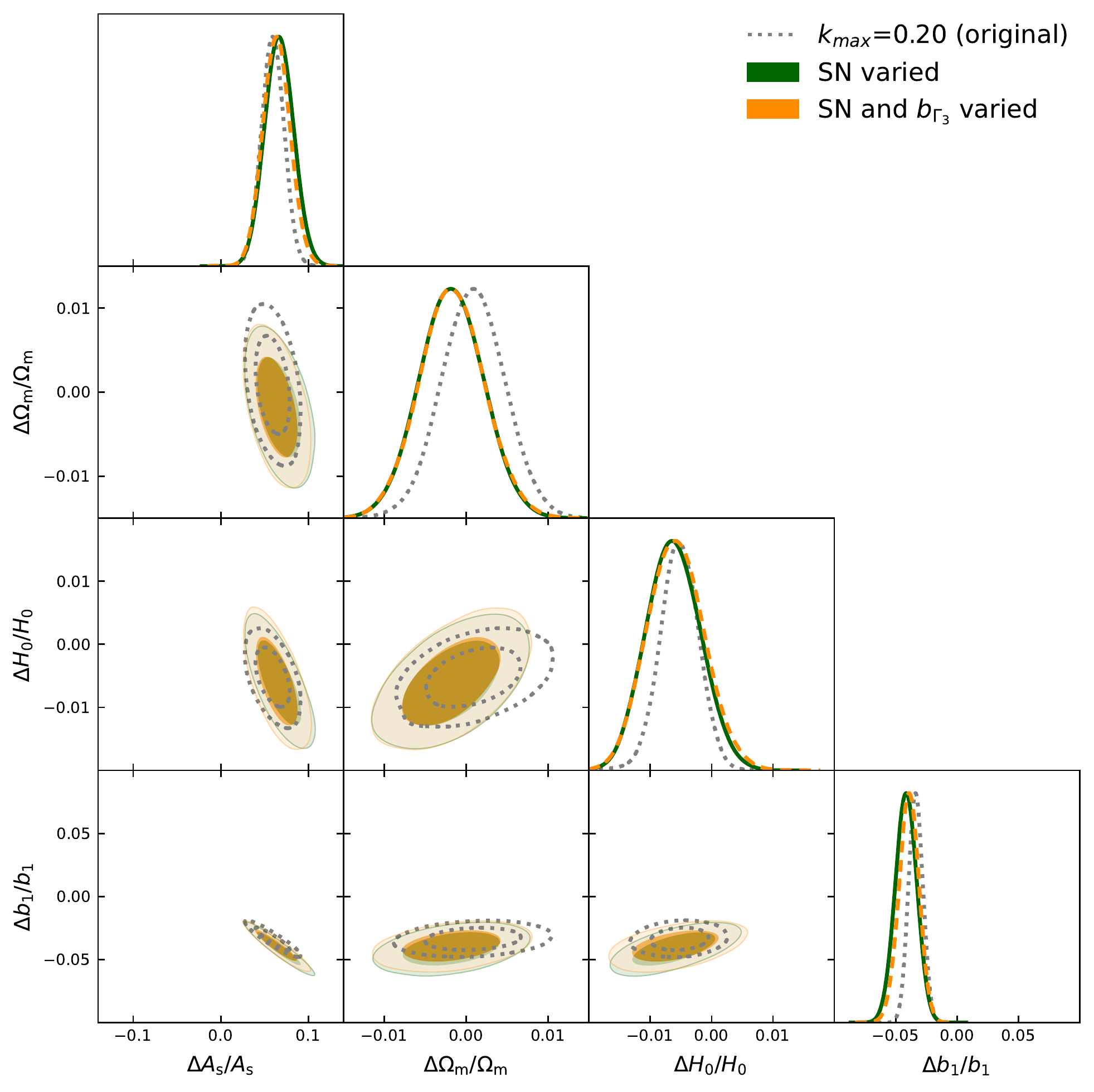}
\end{center}
\caption{
Posterior distributions from the post-unblinding analyses where one or two additional bias parameters are floated. The results are from the pipeline by the East Coast Team.
}
\label{fig:Epost}
\end{figure*}

\begin{figure*}[ht]
\begin{center}
\includegraphics[width=0.49\textwidth]{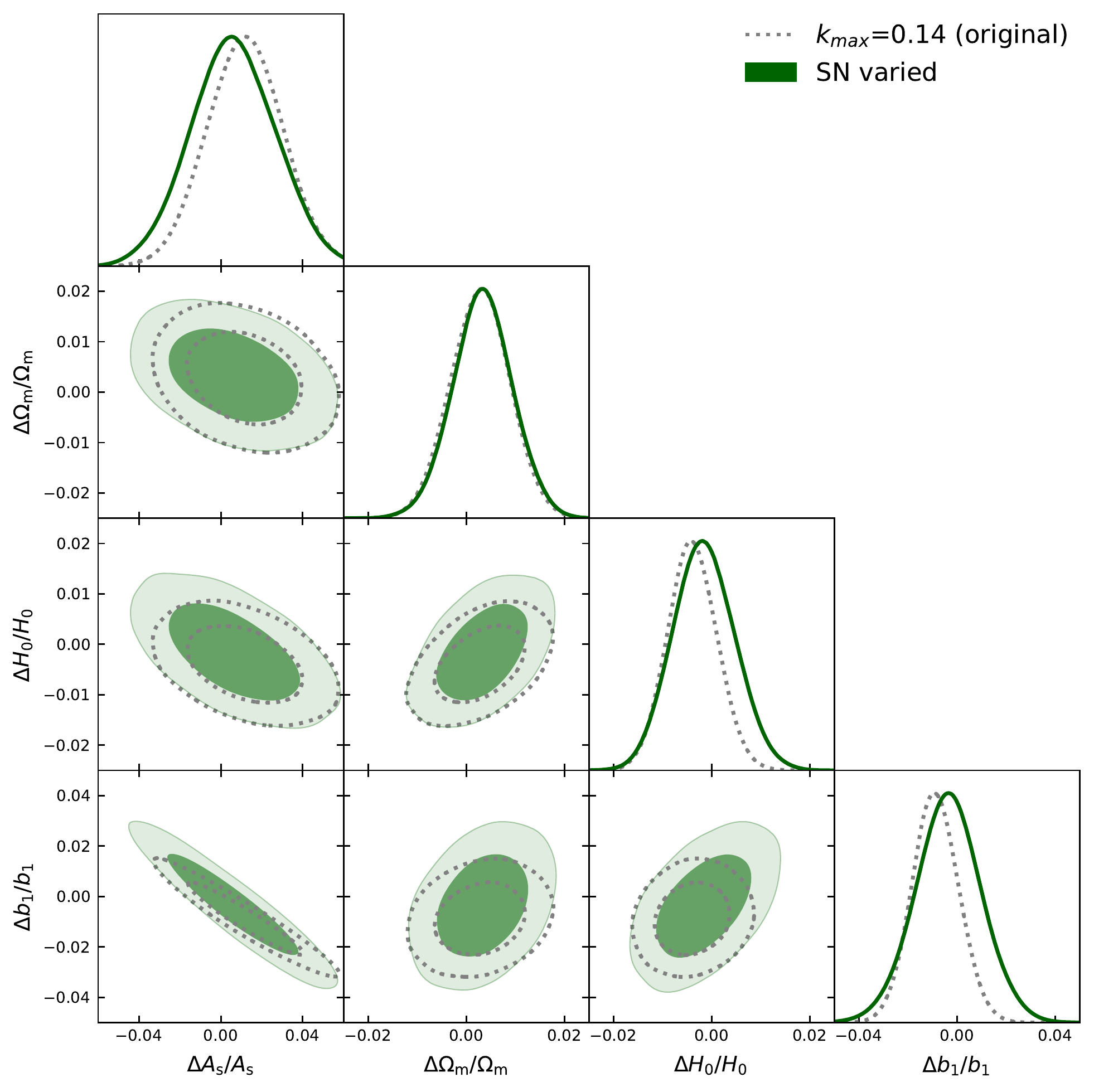}
\includegraphics[width=0.49\textwidth]{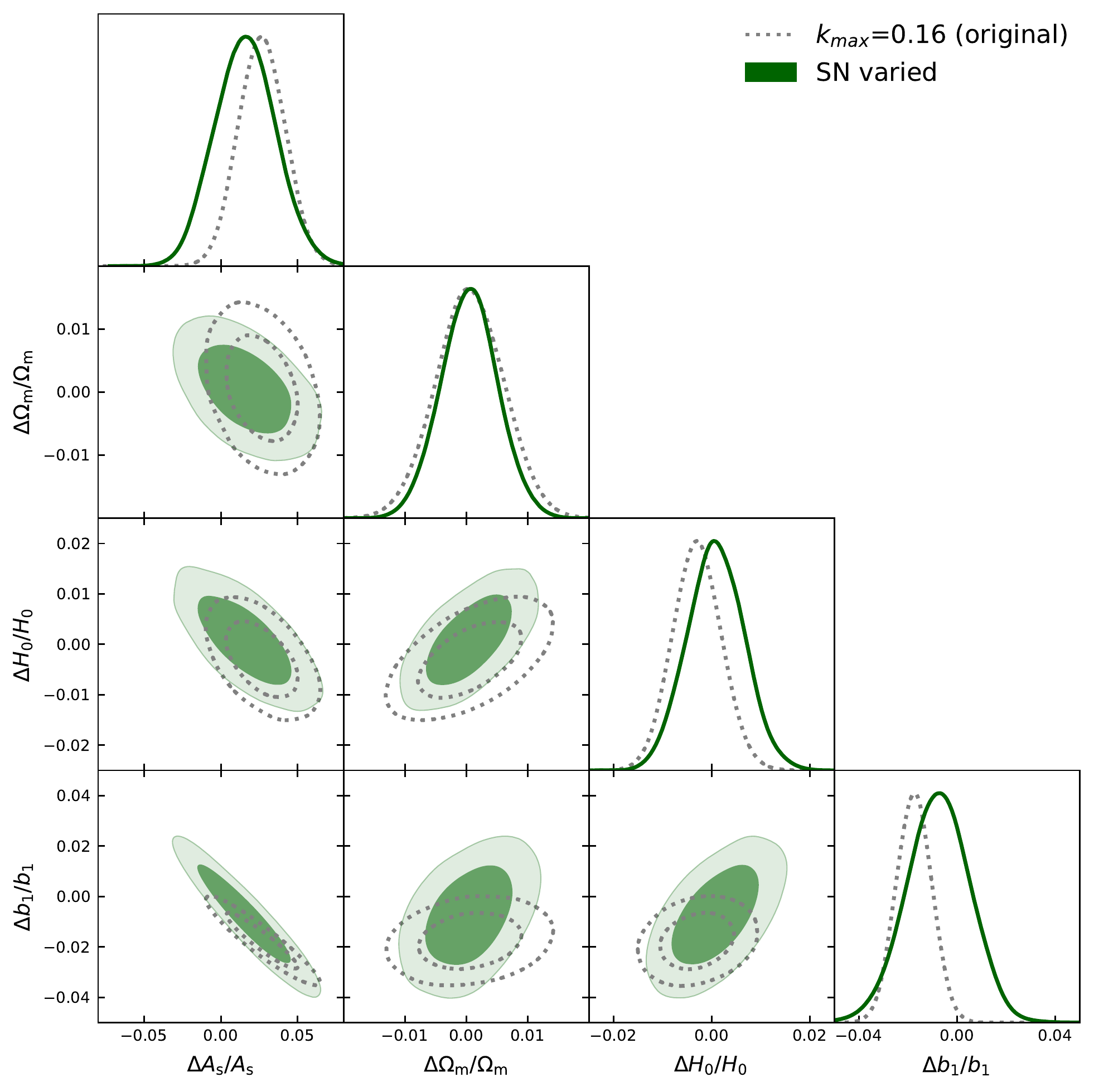}
\includegraphics[width=0.49\textwidth]{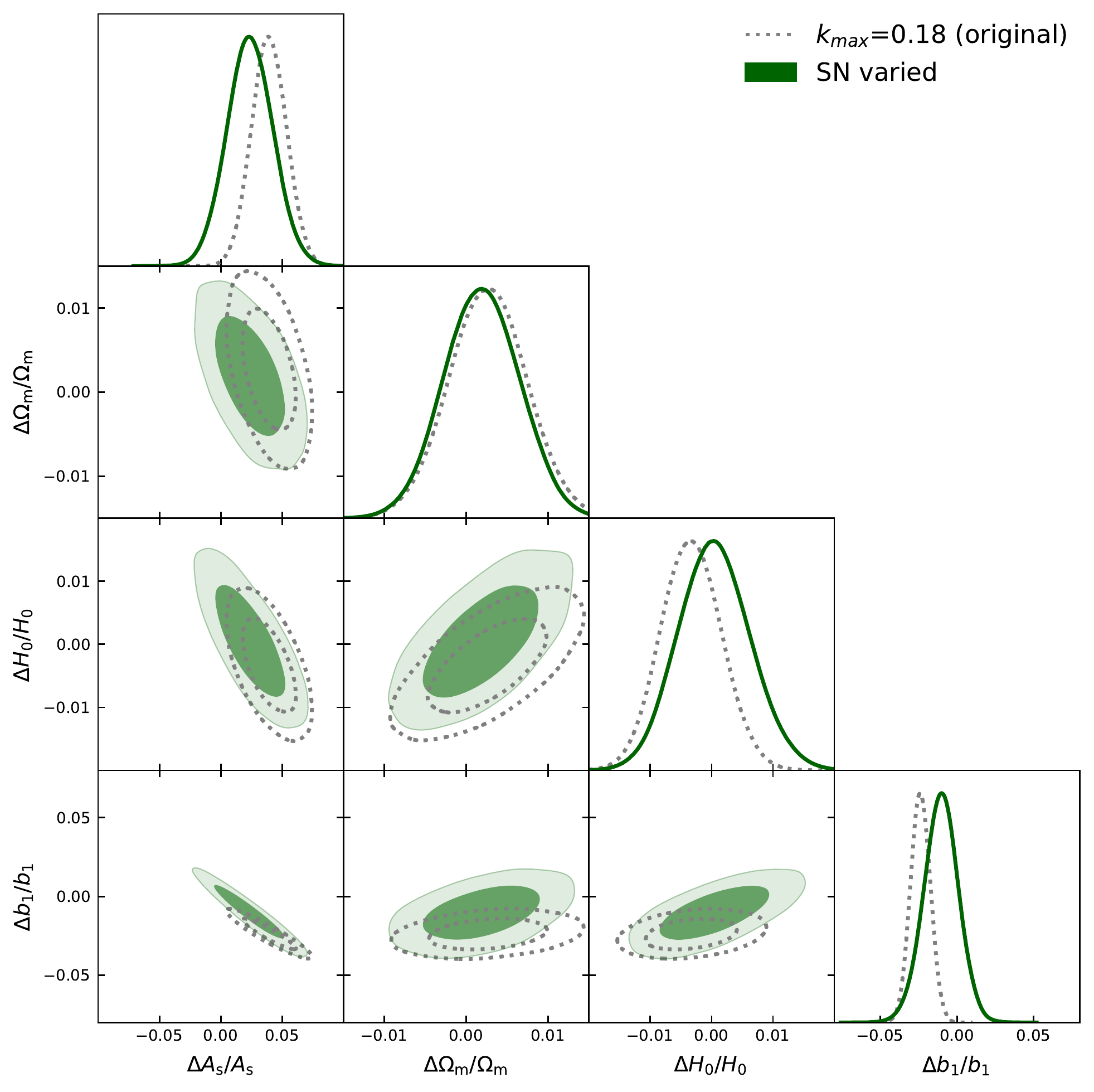}
\includegraphics[width=0.49\textwidth]{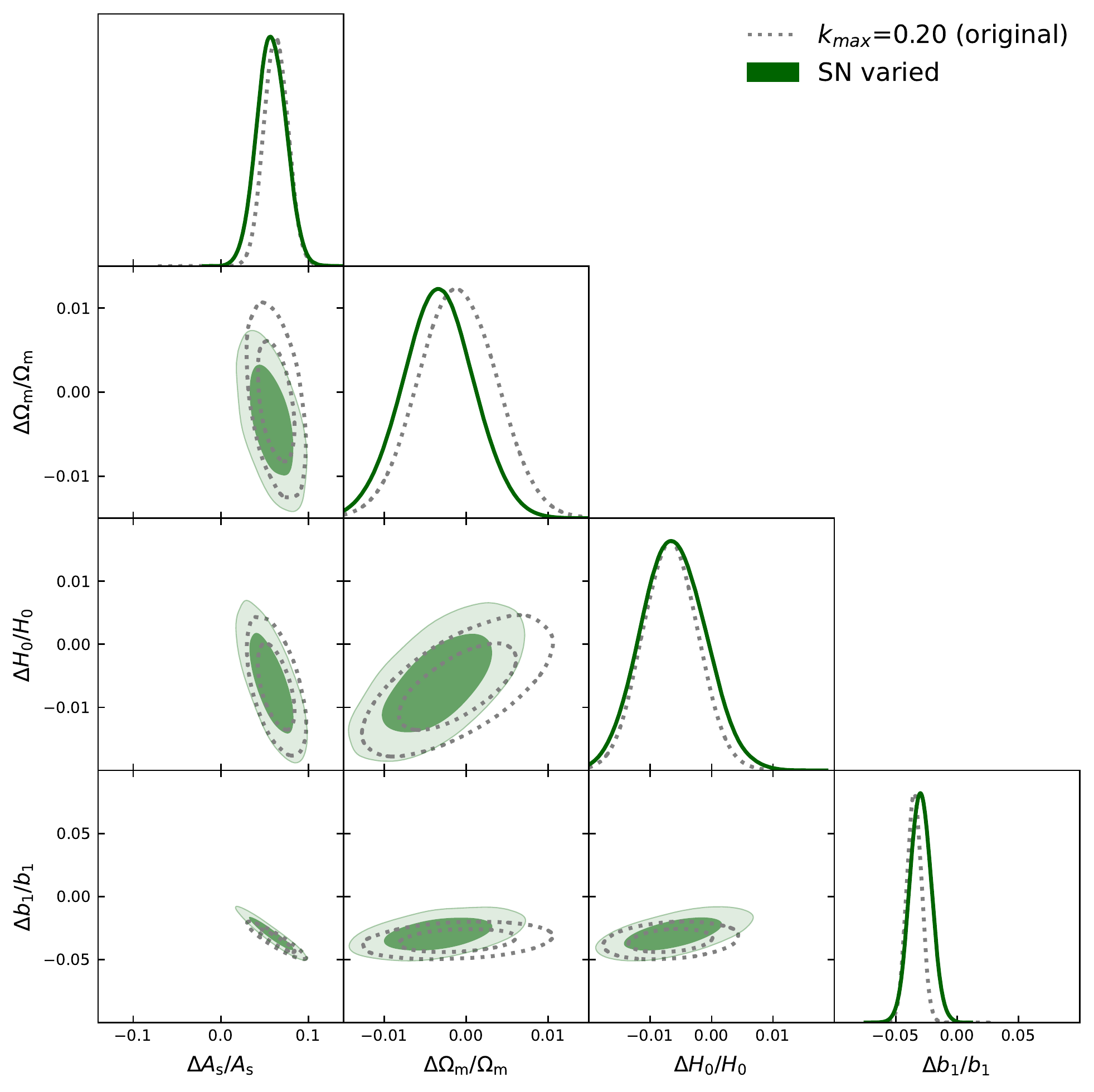}
\end{center}
\caption{
Similar to Fig.~\ref{fig:Epost}, but from the pipeline by the West Coast Team for the case of varied residual shot noise term.
}
\label{fig:Wpost}
\end{figure*}
\subsection{Residual shot noise}
It is known that dark matter halos or associated galaxies are not a Poisson sample of the underlying hypothetical continuous distribution \citep[e.g.,][]{Smith_2007,seljak09}. As explained in Sec.~\ref{subsec:measurement}, the standard shot noise contribution is already subtracted in the power spectra data files provided by the Japan Team. The subtracted shot noise contribution is, strictly speaking, not really an estimate of the additional fluctuations associated with the connection between the underlying smooth field and the discrete point distribution, but simply the ``zero-lag'' correlator inherent in a point process. Therefore, the assumption of the zero shot-noise like term adopted in the blinded analyses presented in the main text is not guaranteed to be valid.
We study here the impact of adding a nuisance parameter to model the residual shot term, which is relevant for the monopole moment.

The green contours in Fig.~\ref{fig:Epost} show the result from the East Coast Team at four different $\kmax$ as indicated in the figure legend.
Fig.~\ref{fig:Wpost} shows the same analysis done by the West Coast Team. Both results are compared with the open dotted contours from the blinded analysis.
Introducing one more free parameter indeed results in  slightly looser constraints with tilted degeneracy directions. An interesting observation is that the biases that the East Coast team displays on $H_0$ and $A_\mathrm{s}$ at $\kmax=0.16$ and $0.18\,\hMpci$ are reduced.
Similarly, the West Coast Team had a bias just on $A_\mathrm{s}$, which is also reduced.
Although the introduction of the non-Poissonian shot-noise term ceases to mitigate the parameter bias at $\kmax=0.2\,\hMpci$ probably due to the absence of terms higher than one-loop in the theoretical template, this parameter would allow for a more robust analysis
in an actual analysis of observational data.
This is exactly what was done by both teams in their analyses of the BOSS data.

\subsection{Floating $b_{\Gamma_3}$}

The East coast team did not vary $b_{\Gamma_3}$ in their analysis because they have found that it does not affect the parameter constraints. This is explicitly illustrated in this appendix.
$b_{\Gamma_3}$
has very little impact
on the power spectrum constraints
because this bias parameter is strongly degenerate with
$b_{\mathcal{G}_2}$.
To break this degeneracy, the East coast team uses the
Gaussian prior centered at the prediction of the coevolution model
\be
b_{\Gamma_3}\sim \mathcal{N}\left(\frac{23}{42}(b_1-1),1\right)
\ee
The results of the analysis with this prior are shown
in Fig.~\ref{fig:Epost}.
One clearly sees that varying $b_{\Gamma_3}$ or fixing it to a constant value has no noticeable impact on the cosmological parameters
and the linear bias $b_1$.

\section{Scaling to realistic surveys}
\label{app:desi}
We have focused on the statistical inference from mock spectra measured from unrealistically large total volume in the main text. To further gain insights to more realistic observations, we redo the analysis after unblinding with much smaller volume. We consider a hypothetical survey with the volume of $25.5\,(\hiGpc)^3$ and scale the error bars from the mock simulations according to the volume ratio. This is close to the expected survey volume of the DESI survey. The results of this appendix should be taken with care, since the effective redshift of the DESI survey is higher than the one we use here, which is similar to the one of the BOSS survey instead. As a consequence, the effect of the nonlinear corrections, and so the systematic error measured here, is larger than what we expect for the actual DESI survey.

\begin{figure*}[ht]
\begin{center}
\includegraphics[width=0.49\textwidth]{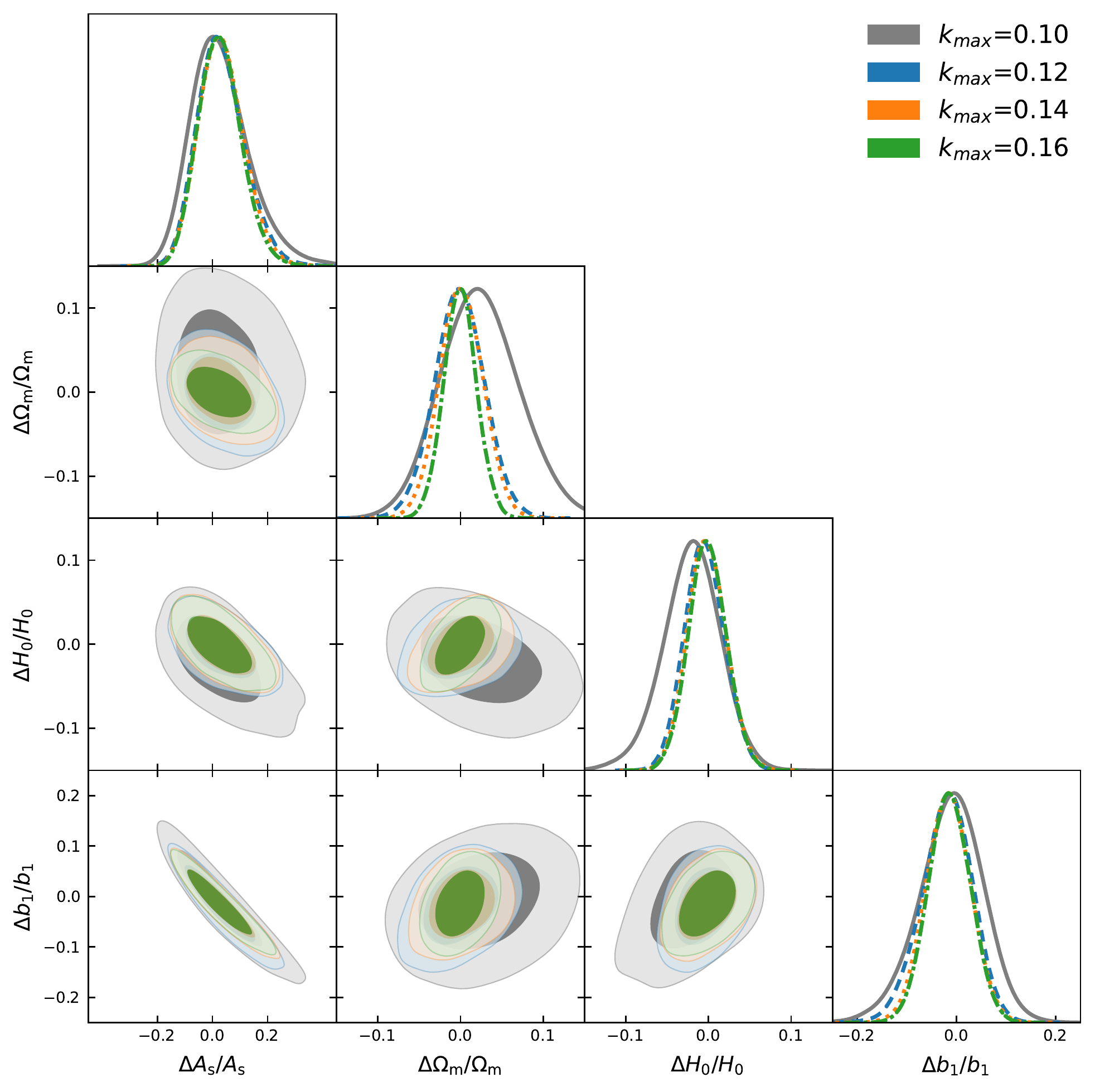}
\includegraphics[width=0.49\textwidth]{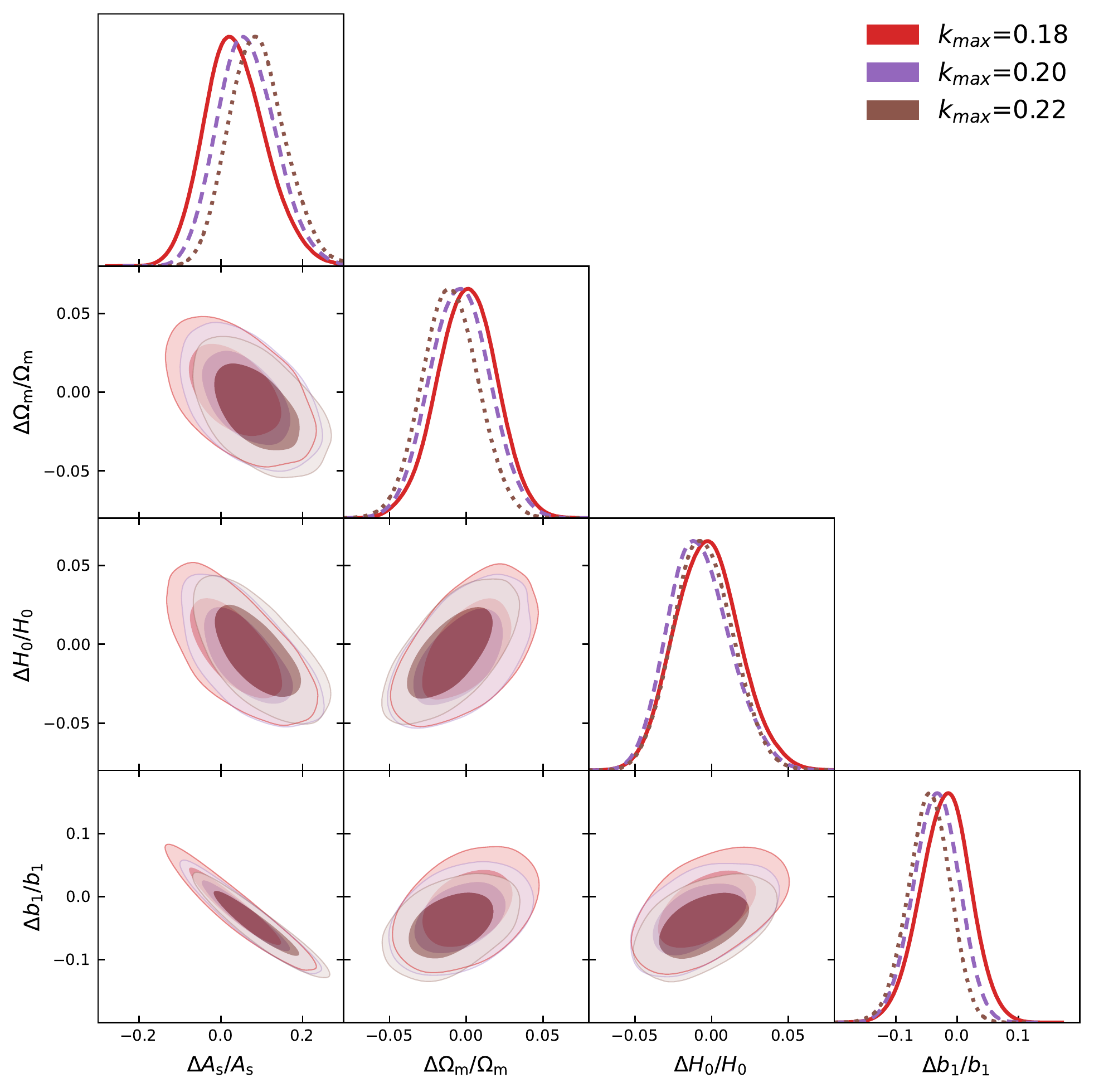}
\end{center}
\caption{
Expected constraints from a DESI-like survey at different maximum wavenumbers (left: $0.1$, $0.12$, $0.14$ and $0.16\,\hMpci$, right: $0.18$, $0.2$ and $0.22\,\hMpci$). We use the pipeline by the West Coast Team used in the main text for this plot.
}
\label{fig:desi1}
\end{figure*}

We show the results using the same analysis pipeline as the one used by the West Coast Team in the main text. Due to the larger error bars in the mock spectra, we can push to smaller scales without bias in the inferred cosmological parameters. Figure~\ref{fig:desi1} shows the 1 and 2-$\sigma$ credible regions for the three varied cosmological parameters as well as the linear bias parameter from this analysis. We show the results at different values of $\kmax$ up to $0.22\hMpci$ as indicated by the figure legend in two panels for ease of visibility (lower $\kmax$ in the left and higher $\kmax$ in the right panel).

Overall, we can see that the inferred parameters are unbiased compared to the statistical error level expected from a DESI-like survey for all the $\kmax$ values considered here.
With respect to the results of the BOSS analysis performed in~\cite{DAmico:2019fhj}, the error bars at $\kmax = 0.2 \hiMpc$ shrink from $3.2\%$ to $2\%$ on $\Omega_m$, from $3.2\%$ to $2.1\%$ on $H_0$ and from $13\%$ to $6.7\%$ on $A_s$.
However, from the runs described in the main text, there is a systematic error on $A_s$ of $\sim 4\%$ at $k_{\rm max} = 0.2 \hiMpc$, which corresponds to about $2/3$ of the statistical error here.
These results are particularly encouraging in view of the fact that DESI will survey a higher redshift than the one of the simulations, where nonlinear corrections will be less important.

\bibliography{lssref}
\bibliographystyle{apsrev}

\end{document}